\documentclass[iop]{emulateapj}

\def\lapp{\ifmmode\stackrel{<}{_{\sim}}\else$\stackrel{<}{_{\sim}}$\fi}
\def\gapp{\ifmmode\stackrel{>}{_{\sim}}\else$\stackrel{>}{_{\sim}}$\fi}
\usepackage{multirow}
\usepackage{color}
\usepackage{amsmath}
\usepackage{soul}
\usepackage{hyperref}

\begin{document}

\title{High-Energy Variability of PSR~J1311$-$3430}

\author{
Hongjun An\altaffilmark{1,2}, Roger W. Romani\altaffilmark{2},
Tyrel Johnson\altaffilmark{3}, 
Matthew Kerr\altaffilmark{3}, and
Colin J. Clark\altaffilmark{4,5}
\\
{\small $^1$Department of Astronomy and Space Science, Chungbuk National University, Cheongju, 28644, Republic of Korea}\\
{\small $^2$Department of Physics/KIPAC, Stanford University, Stanford, CA 94305-4060, USA}\\
{\small $^3$College of Science, George Mason University, Fairfax, VA 22030, USA}\\
{\small $^4$Albert-Einstein-Institut, Max-Planck-Institut f{\"u}r Gravitationsphysik, D-30167 Hannover, Germany}\\
{\small $^5$Leibniz Universit{\"a}t Hannover, D-30167 Hannover, Germany}\\
}
\altaffiliation{$^2$hjan@chungbuk.ac.kr}

\begin{abstract}
	We have studied the variability of the black-widow type binary millisecond pulsar
PSR~J1311$-$3430 from optical to gamma-ray energies. We confirm evidence for
orbital modulation in the weak off-pulse $\gapp$200-MeV emission, with a peak at
$\phi_B\approx 0.8$, following pulsar inferior conjunction. The peak has a relatively hard
spectrum, extending above $\sim 1$\,GeV.
{\it XMM-Newton} and {\it Swift} UV observations also show that this source's strong
X-ray flaring activity is associated with optical/UV flares. 
With a duty cycle $\sim 7-19$\%, this flaring is quite prominent with an 
apparent power-law intensity distribution. Flares are present at all 
orbital phases, with a slight preference for $\phi_B=0.5-0.7$.  We explore possible 
connections of these variabilities with
the intrabinary shock and magnetic activity on the low mass secondary.
\end{abstract}

\keywords{{binaries: close --- gamma rays: stars --- X-rays: binaries
--- pulsars: individual (PSR~J1311$-$3430)}}

\section{Introduction}

        ``Black Widow''s (BWs) are energetic millisecond pulsars (MSPs) in short-period
binaries driving a strong evaporative wind from their low-mass companion star. Optical
observations of these companions often show the effect of this heating but in many cases
the light curves do not match those expected from direct illumination by the pulsar
\citep[e.g. PSR~J2215$+$5135;][]{sh14,rgf15}. Instead, the heating power may be reprocessed
in an intrabinary shock (IBS) set up between the relativistic pulsar wind and the massive
wind driven from the companion \citep[][]{rs16}. Thus the properties of this shock and its
interaction with the companion star will be critical to understanding the evolution
of BW MSPs.

        Happily we see evidence for the wind in a range of other wavebands. In particular
strong and variable radio eclipses of evaporating pulsars testify to an ionized plasma
in the outflow. In a few cases \citep[e.g. PSR~J1311$-$3430;][]{rfb15} we can see optical
emission lines from atoms in the companion wind. At higher energies the relativistic shocked plasma
in the IBS can contribute to strong orbital modulation in the X-ray band \citep{grmc+14},
with caustic peaks due to beamed, likely synchrotron, emission. This high-energy emission is
particularly valuable in helping map the geometry of the IBS. In addition, the variability
of these systems from brief flares to the episodic accretion state changes of the so-called
``transitioning'' MSPs, such as PSR~J1023$+$0038 \citep{asrk+09},  gives important clues
to the dynamics of the IBS.

        We report here on the high-energy variability of an extreme member of
this class, PSR~J1311$-$3430 (hereafter J1311). With its $P_b=1.56$\,hr orbital period
\citep{r12} the shortest among confirmed MSPs \citep{pgf+12} and a $\sim 0.01M_\odot$
He-rich companion indicating membership in the extreme ``Tidarren'' sub-class \citep[][]{rgfz16},
this system has a particularly powerful IBS. It is also interesting as a possible
high-mass neutron star and as a likely descendant of an Ultra-Compact X-ray Binary
(UCXB) and progenitor of an isolated MSP. J1311 shows rich temporal variability,
including claimed orbital modulation in off-pulse gamma rays \citep{xw15} and rapid optical
flares \citep[][]{rfb15}. Here we study the orbital and short-period variabilities of J1311
for insights into the IBS physics and the companion heating, using
{\it Fermi} Large Area Telescope \citep[LAT;][]{fermimission} data and archival optical
and X-ray observations. In section~\ref{sec:sec2}, we describe the observational data
we use. We show the analysis results in section~\ref{sec:sec3}. We then discuss
and conclude in section~\ref{sec:sec4}.

\section{Observational Data Sets and Basic Processing}
\label{sec:sec2}
        We use $\sim$8 years of {\it Fermi} LAT \citep{fermimission} data collected between
2008 August 04 and 2016 May 02 UTC, downloaded from the Pass-8 \citep[][]{fermiP8} event
archive and further analyzed with the {\it Fermi}-LAT Science Tools {\tt v10r0p5}
along with P8R2\_V6 instrument response functions.
For our analyses, we use source class events with Front/Back event type in an
$R=10^\circ$ aperture. We analyzed the 100\,MeV--300\,GeV and the 100\,MeV--500\,GeV data
for the spectral and the timing analyses, respectively.
We further employed standard $<90^\circ$ zenith angle and $52^\circ$ rocking angle cuts.

        For X-ray data, we use archival {\it Swift}, {\it Suzaku}, {\it XMM-Newton}
and {\it Chandra}
observations. The {\it Swift} observations were made between MJD~54889 and 56731 and
comprised 47 short observations. The total net {\it Swift} exposure is 90\,ks.
We processed the data using the XRT standard pipeline tools integrated with Heasoft~6.16 along
with the HEASARC remote CALDB.
Two {\it Suzaku} observations were taken in MJD~55047 (AO4, obs. ID 804018010, 40\,ks) and
MJD~55774 (AO6, obs. ID 706001010, 80\,ks). We reprocessed the XIS data using the {\tt aepipeline}
tool in Heasoft~6.16.
Results for analyses of these data have been
reported previously \citep[][]{r12,kyk+12}. The {\it XMM-Newton} data were taken on
MJD~56871 for 130\,ks (obs. ID 0744210101). The EPIC data were processed using the
{\tt epproc} and the {\tt emproc} tools in SAS version 20141104\_1833.
The archival {\it Chandra} data \citep[][]{cdgs+12,apg15} have exposures of 20\,ks (Obs. ID 11790)
and 10\,ks (Obs. ID 13285), and are processed with {\tt chandra\_repro} in CIAO~4.7.2 to use the
most recent calibration files.

        For optical data, we use the archival {\it Swift} UVOT and {\it XMM-Newton} optical
monitor observations. We processed these data using the {\tt uvotsource} tool and the
{\tt omichain} tool, respectively. These were compared with ground-based
light-curve studies \citep[][]{rfb15}.
	
\section{Data Analysis and Results}
\label{sec:sec3}
\subsection{Fermi-LAT data analysis}
\label{sec:sec3_1}

       We first performed a gamma-ray spectral analysis.
We used a $R=10^\circ$ region of interest (RoI), and analyzed the data
following the binned likelihood analysis as described in
{\it Fermi} Science Support
Center (FSSC)\footnote{https://fermi.gsfc.nasa.gov/ssc/data/analysis/documentation\\/Pass8\_usage.html}.
When fitting the data, we used a power-law
with exponential cutoff model (PLEXP),
$dN/dE=N_0(E/E_0)^{-\Gamma_{\rm 1}}\mathrm{exp}(-(E/E_{\rm c})^{\Gamma_{\rm 2}})$,
with $\Gamma_{\rm 2}$ held fixed at 1 for J1311
and fit all the bright sources ($\gapp$5$\sigma$) in the aperture,
the Galactic diffuse emission \citep[{\tt gll\_iem\_v06.fits};][]{fermigllv06},
and the isotropic diffuse emission \citep[{\tt iso\_P8R2\_SOURCE\_V6\_v06.txt};][]{fermiiso}
in the 100\,MeV--300\,GeV band with {\tt pylikelihood}. In the fit, we also include
all the 3FGL sources within 20$^\circ$ \citep{fermi3fgl}.
We consider energy dispersion in the analysis\footnote{https://fermi.gsfc.nasa.gov/ssc/data/analysis/scitools/binned\\\_likelihood\_tutorial.html and https://fermi.gsfc.nasa.gov/ssc/data\\/analysis/documentation/Pass8\_edisp\_usage.html},
and therefore processed the data in the wider 60\,MeV--500\,GeV band.
The best-fit parameters for J1311 are $\Gamma_{\rm 1}=1.87\pm0.03\pm0.04$, $E_{\rm c}=5.3\pm0.5\pm0.5$\,GeV, with
100\,MeV--300\,GeV flux of $9.2\pm0.3\pm0.2\times10^{-8}\rm \ phs\ cm^{-2}\ s^{-1}$ (see Table~\ref{ta:ta1}),
consistent with the previous LAT spectral analysis within the uncertainties
\citep{fermi3fgl}.
The first and second uncertainties given for each parameter represent 1-$\sigma$
statistical and systematic uncertainties, respectively.
Note that the systematic uncertainties are estimated by varying the scale for the
response functions\footnote{https://fermi.gsfc.nasa.gov/ssc/data/analysis/scitools/Aeff\_S\\ystematics.html\#bracketing}
and normalization of the interstellar diffuse model (gll\_iem\_v06) by 6\%.
We then calculated changes in the best-fit parameters due to each systematic components
and summed the errors in quadrature.
We also varied the model; we hold all the parameters except
for those of J1311 fixed at the 3FGL values or varied the number of sources to fit.
We find that the results are all
consistent with one another within the uncertainties. We further tried different
RoI's, $R=5^\circ$ and $R=15^\circ$, and performed similar studies. The results for these
new RoI's are also consistent with those for $R=10^\circ$ within the uncertainties.

\begin{figure}
\centering
\hspace{-3.0 mm}
\includegraphics[width=3.3 in]{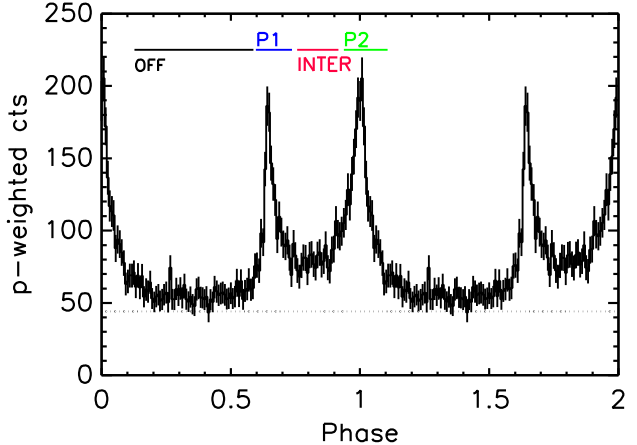}
\figcaption{Gamma-ray pulse profile (128 bins) in the 100\,MeV--500\,GeV and $R<2^\circ$.
See section~\ref{sec:sec2} for the timing solution.
The phase-averaged background level estimated following \citet{fermi2pc} is shown with a dashed line.
\label{fig:fig1}
}
\vspace{0mm}
\end{figure}

\newcommand{\markaa}{\tablenotemark{a}}
\newcommand{\markbb}{\tablenotemark{b}}
\begin{table}[t]
\vspace{-0.0in}
\begin{center}
\caption{Fit results for the orbital-phase averaged and 
orbital-phase revolved {\it Fermi}-LAT data in the ``OFF'' pulse-phase interval
}
\label{ta:ta1}
\vspace{-0.05in}
\scriptsize{
\begin{tabular}{ccccc} \hline\hline
\multicolumn{4}{c}{Spin-phase resolved and orbital-phase averaged} \\
Phase	& $\Gamma_{\rm 1}$	& $E_{\rm c} $	&$F$\markaa  \\ 
		& 			& (GeV) 		&	   \\ \hline
ALL  		& $1.87\pm0.03\pm0.04$	& $5.3\pm0.5\pm0.5$	& $9.2\pm0.3\pm0.2$	 \\
P1 & $1.71\pm0.05\pm0.06$ 	& $4.7\pm0.7\pm0.4$	& $14.5\pm0.6\pm1.4$	 \\
Inter & $1.67\pm0.07\pm0.11$	& $3.7\pm0.7\pm0.6$	& $8.1\pm0.5\pm1.3$	 \\
P2 & $1.80\pm0.04\pm0.05$	& $5.4\pm0.8\pm0.4$	& $18.6\pm0.6\pm1.6$	 \\
OFF& $2.24\pm0.12\pm0.20$	& $2.0\pm0.7\pm0.4$	& $4.5\pm0.3\pm1.3$      \\ \hline \hline
\multicolumn{4}{c}{Orbital-phase resolved in the ``OFF'' pulse-phase interval} \\ 
Dip  & $2.54\pm0.11\pm0.11$	& $3.3\pm0.0$\markbb	& $4.3\pm0.5\pm1.2$      \\ 
Hump & $2.33\pm0.13\pm0.19$	& $3.3\pm1.7\pm0.6$	& $5.2\pm0.4\pm1.3$	 \\ \hline
\end{tabular}}
\end{center}
\vspace{-0.5 mm}
\footnotesize{
{\bf Notes.} 1-$\sigma$ statistical and systematic uncertainties are shown.\\}
$^{\rm a}${100\,MeV--300\,GeV flux in units of $10^{-8}\ phs\rm \ cm^{-2}\ s^{-1} $.}\\
$^{\rm b}${Held fixed at the best-fit value for the Hump spectrum (see Figure~\ref{fig:fig2}).}\\
\end{table}

        To isolate the strong gamma-ray pulsations, we checked the timing
solution of \citet{pgf+12}, folding the {\it Fermi} LAT events.
These events were selected from small apertures ($R$$\lapp$$2^\circ$). The source
probabilities of each event, based on energy and position, were computed  using {\tt gtsrcprob},
and the time series was phased and folded using {\tt tempo2}. Although the pulsar spin
modulation shows very strongly, this solution, based on earlier LAT data, leaves a large
drift in the pulse arrival time residual especially after MJD~56600.
We therefore derived a new solution covering all data until MJD 57510.

	Starting from the timing solution of \citet{pgf+12}, we used the PINT software
package\footnote{{https://github.com/nanograv/PINT}} and an unbinned Markov chain
Monte Carlo maximum likelihood sampling technique \citep{TucksThesis}.
We modeled the pulse profile using two asymmetric Lorentzians for the main peaks
and a simple Gaussian for the bridge emission.  The parameters of the template
were varied jointly with the timing model parameters, allowing us to marginalize
over the profile model. We found that accurately tracking the pulse phase required
the addition of an orbital period derivative ($\dot P_B$),
proper motion ($\mu_\alpha \mathrm{cos} \delta$ and $\mu_\delta$),
and a second spin frequency derivative ($\ddot{\nu}$). Orbital eccentricity is undetected
and fixed at $e=0$. The resulting best-fit parameters and their 1-$\sigma$ confidence intervals
from the posterior distribution are reported in Table~\ref{ta:ta2},
and the resulting pulse profile appears in Figure~\ref{fig:fig1}.

\newcommand{\markap}{\tablenotemark{a}}
\begin{table}
\centering
\caption{Timing Parameters for PSR~J1311$-$3430.}
\label{ta:ta2}
\begin{tabular}{ c c }
\hline
\hline
\rule{0pt}{3ex}
RA (J2000)              & $13^h11^m45.72358(1)^s$\\
\rule{0pt}{3ex}
DEC (J2000)            & $-34^\circ  30' 30.342(3)''$\\
\rule{0pt}{3ex}
$\mu_\alpha \mathrm{cos} \delta$ (mas/yr)          & $-$6.8(6) \\
\rule{0pt}{3ex}
$\mu_\delta$ (mas/yr)     & $-$3.5(8) \\
\rule{0pt}{3ex}
Position Epoch (MJD)          & 56228 \\
\hline
\rule{0pt}{3ex}
$\nu$ (s$^{-1}$) & 390.56839299885(1) \\
\rule{0pt}{3ex}
$\dot \nu$ (s$^{-2}$) & $-$3.1882(3)$\times 10^{-15}$ \\
$\ddot \nu$ (s$^{-3}$) & 1.03(8)$\times 10^{-25}$ \\
\rule{0pt}{3ex}
Epoch (MJD)             & 56228 \\
\hline
Binary model  & ELL1\markap \\
$P_{\rm B}$  (day) & 0.0651157347(2) \\
$\dot P_{\rm B}$  ($s\ s^{-1}$) & 3.7(3)$\times10^{-12}$ \\
Projected semi-major axis  ($lt$-$s$) & 0.010575(2) \\
$T_{\rm ASC}$ (MJD) & 56009.129455(4) \\
\hline
\hspace{0.5 mm}
\end{tabular}\\
$^{\rm a}${See \citet{ehm06} for the timing model definition.}\\
\end{table}

\begin{figure*}
\centering
\begin{tabular}{ccc}
\hspace{-3.0 mm}
\includegraphics[width=2.3 in]{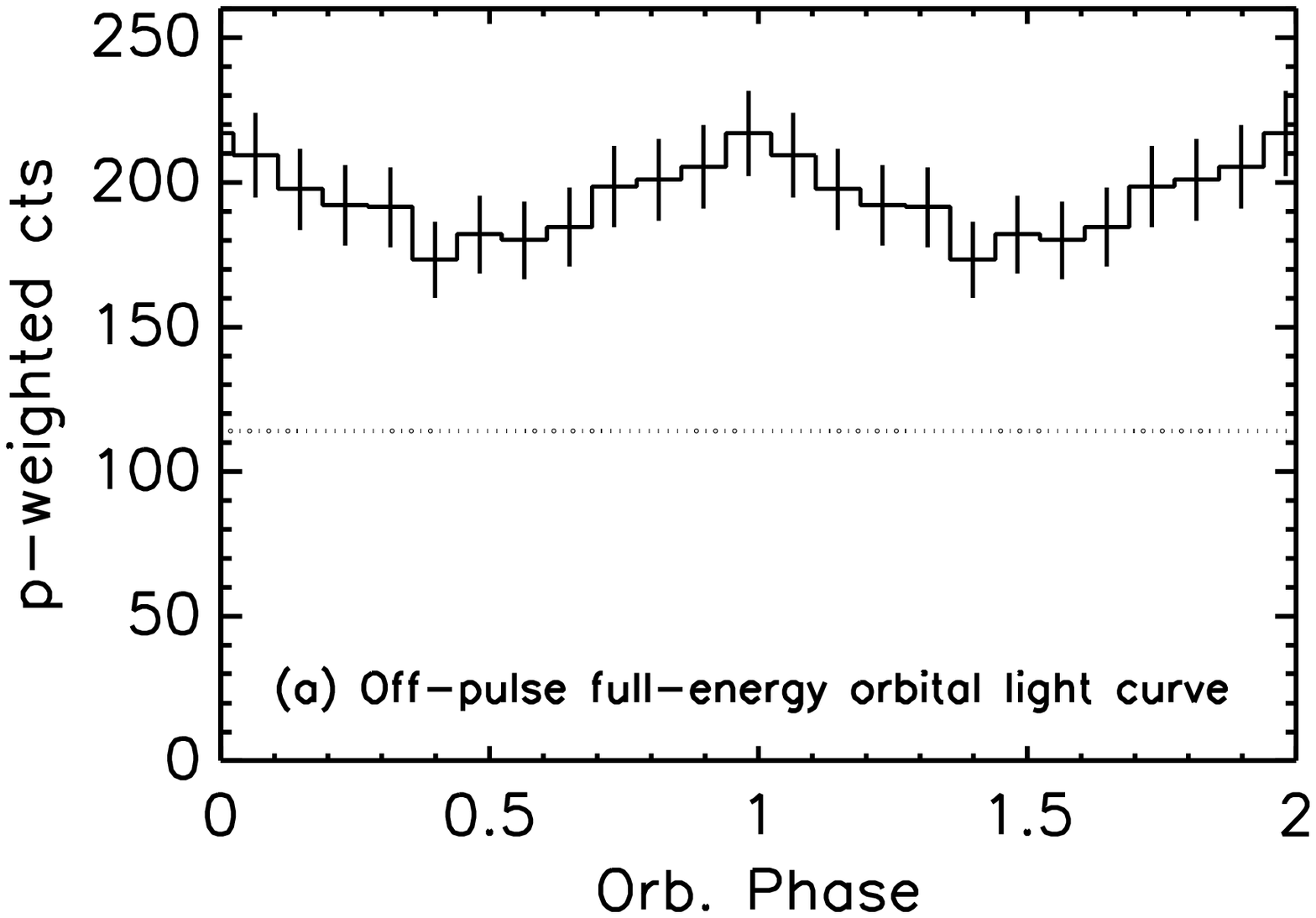} &
\hspace{-3.0 mm}
\includegraphics[width=2.3 in]{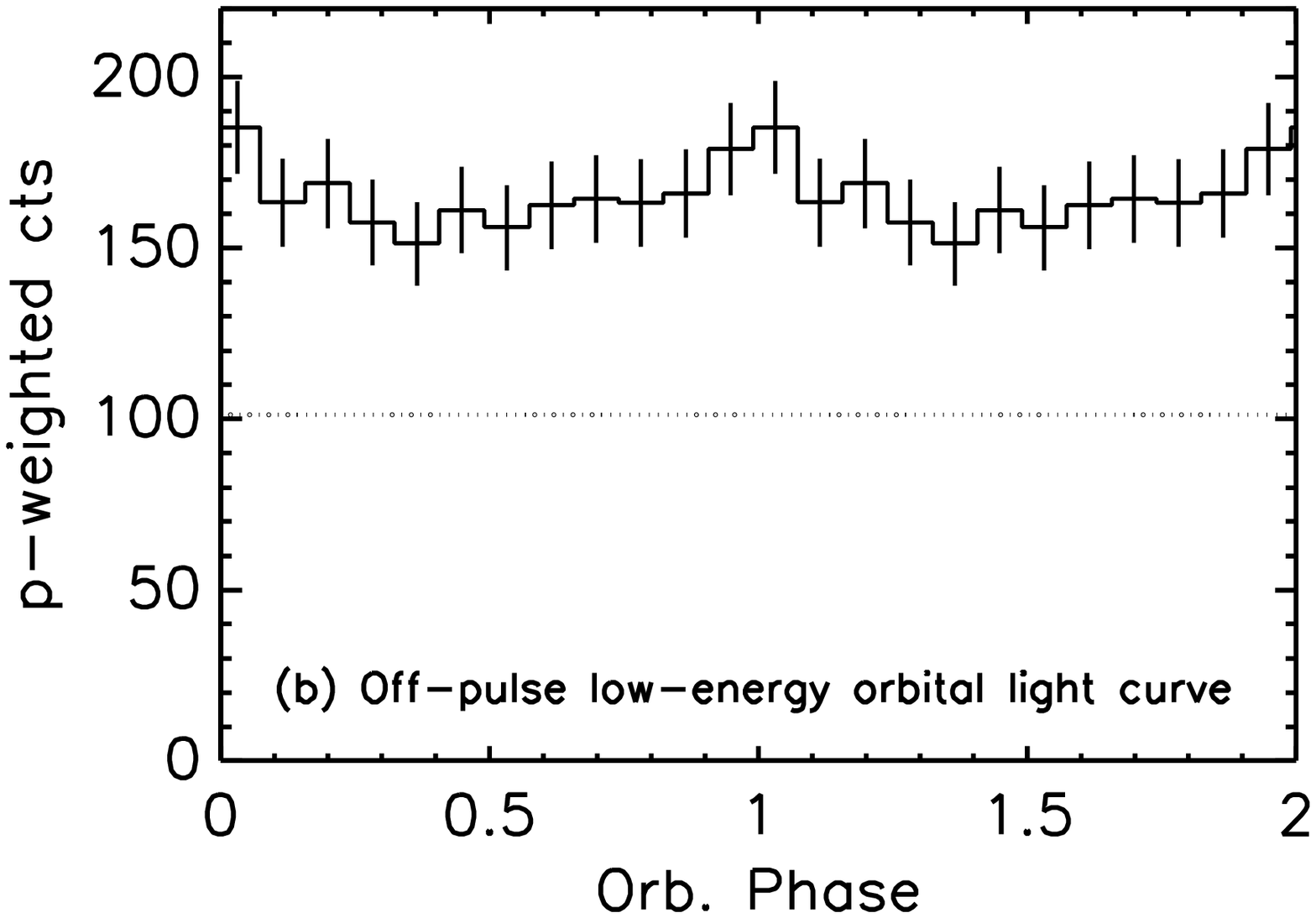} &
\hspace{-3.0 mm}
\includegraphics[width=2.3 in]{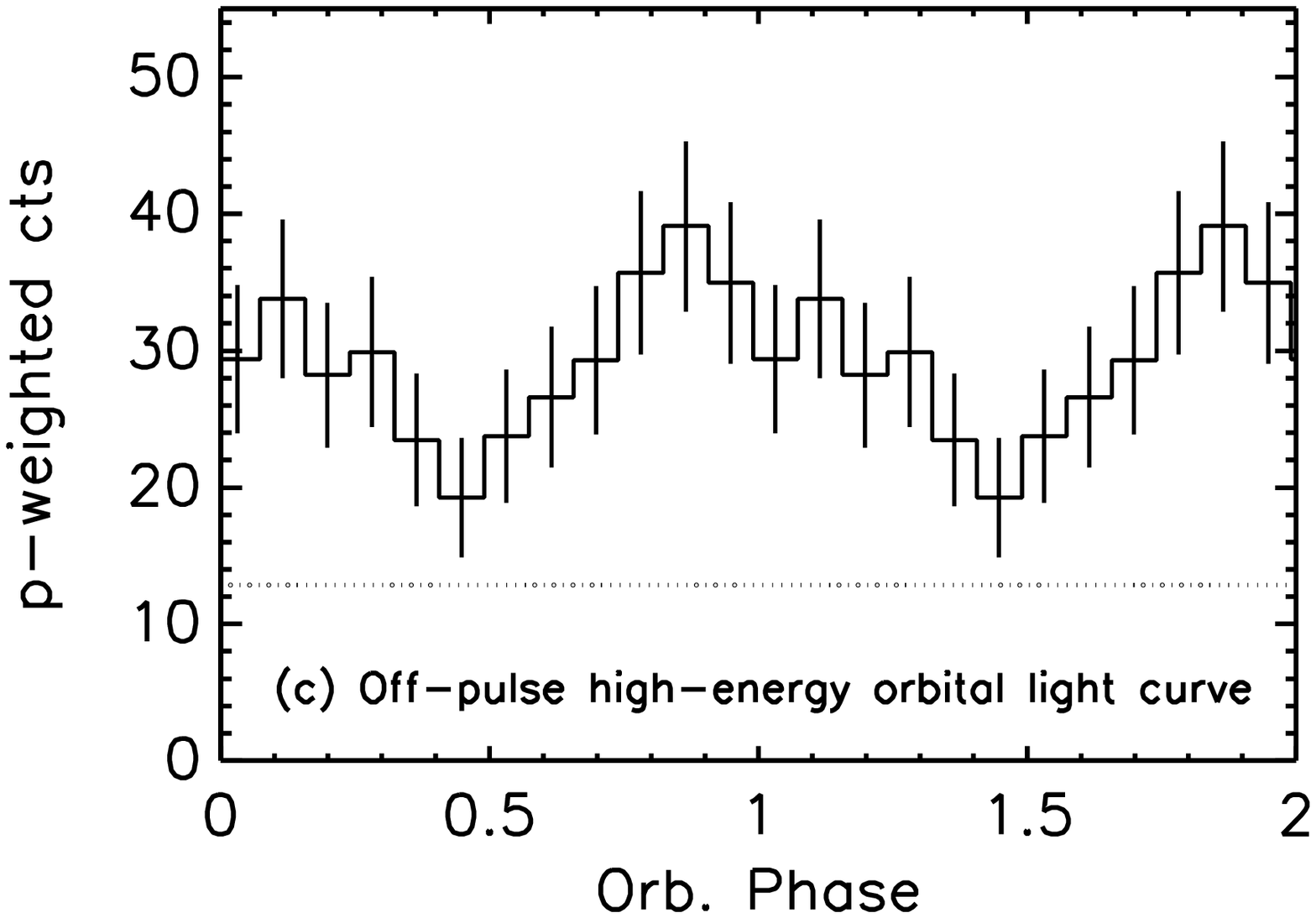} \\
\end{tabular}
\figcaption{Off-pulse phase ($\Delta \phi_P$=0.127--0.587) orbital light curves
taken with a $R<2^\circ$ aperture (MJD~54682--57510) for several different energy bands:
(a) $E$=100\,MeV--100\,GeV, (b) $E$=200\,MeV--1\,GeV, (c) $E$=1\,GeV--100\,GeV.
See Figure~\ref{fig:fig1} for the off-pulse definition and
section~\ref{sec:sec2} for the timing solution.
The phase-averaged background level estimated following \citet{fermi2pc} is shown with a dashed line.
\label{fig:fig2}
}
\vspace{0mm}
\end{figure*}

	By cutting on spin phase ($\phi_P$) to isolate the off-pulse region, one can minimize
the pulsar magnetospheric flux and search for non-magnetospheric emission.
In a study of 2,500 days of Pass-7 LAT data \citet{xw15} found evidence for an
orbital modulation of the off-pulse $>0.2$\,GeV flux,
finding $\sim$3$\sigma$ (0.003 chance probability) significance.
To study such emission with our improved data set we constructed orbital light curves
in the off- ($\phi_P=0.127-0.587$) and three on-pulse
(P1, P2, and the interpulse, Figure~\ref{fig:fig2}) intervals.
A concern for this study is that the orbital period $P_b\simeq 94$\,mins is very close to that
of the {\it Fermi} satellite (96\,mins), hence artificial modulation may be produced due to
the satellite's orbital motion.
We therefore folded time-resolved exposure ($\Delta T=30$\,s) on the orbital period and
confirmed that exposure variations across the phase bins used for
the orbital light curves in Figure~\ref{fig:fig2} were negligible ($\lapp$1\%).
We also confirmed that varying zenith angle
cuts to adjust the imperfect exclusion of Earth-limb gamma rays introduced
no variation beyond the simple scaling due to changes in source count
statistics.

\begin{figure*}
\centering
\begin{tabular}{cc}
\hspace{-7.0 mm}
\includegraphics[width=3.5 in]{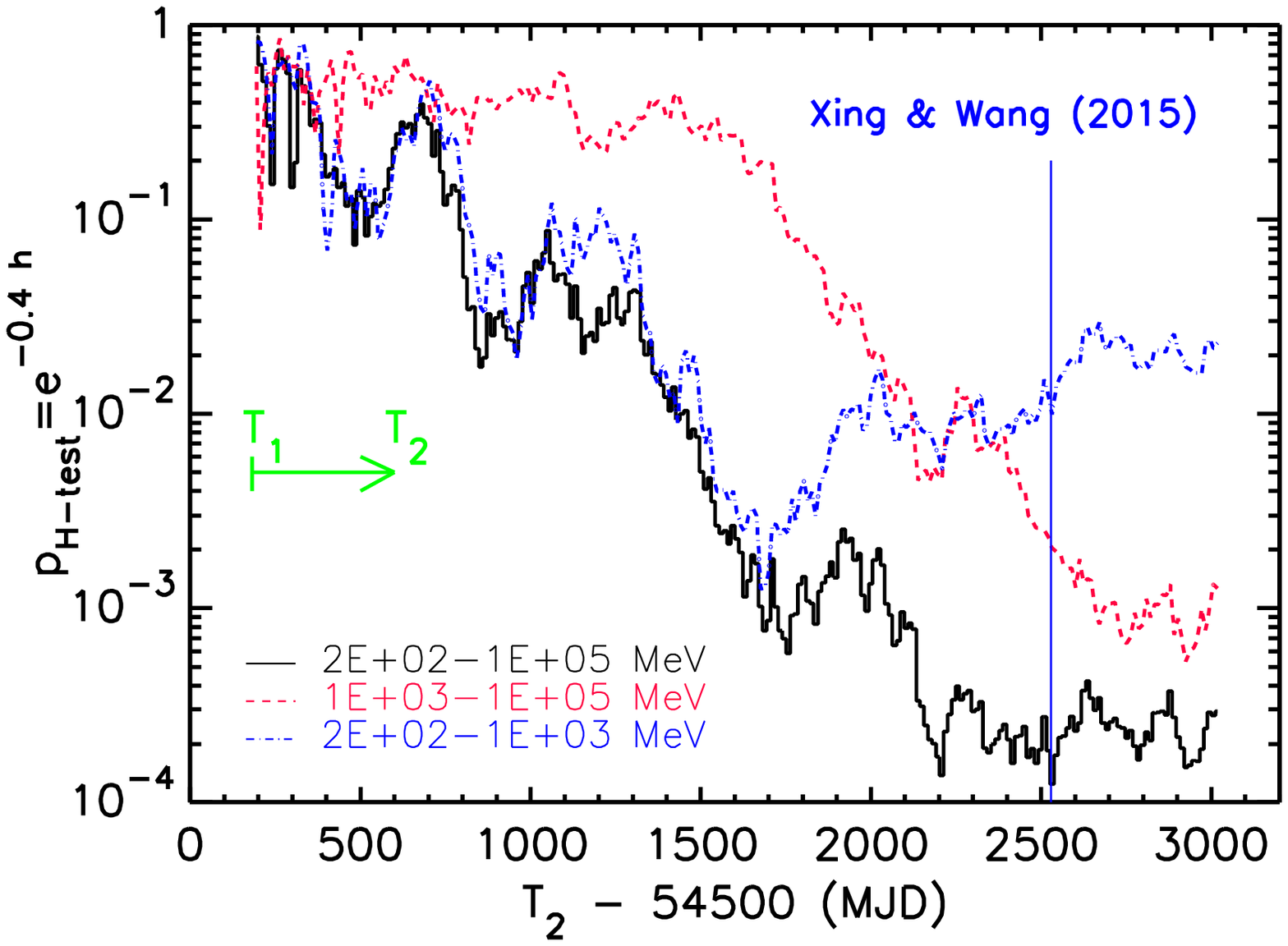} &
\includegraphics[width=3.5 in]{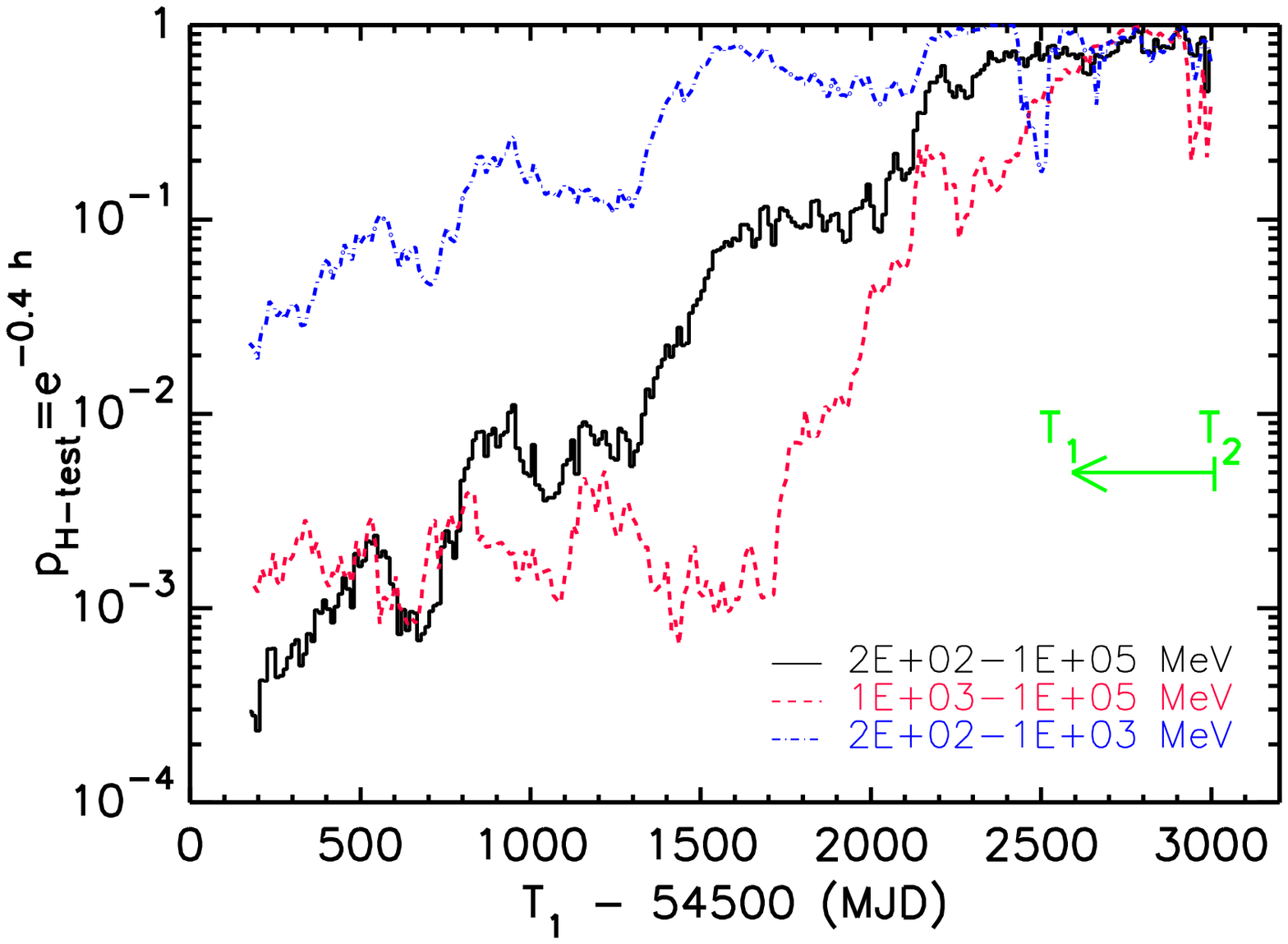} \\
\end{tabular}
\figcaption{Forward (left) and reverse (right) time-cumulative
probabilities for the null hypothesis of a flat orbital light curve from
the weighted counts shown in Figure~\ref{fig:fig2}.
The time intervals for these plots are $T_2 - T_1$, where for fixed
$T_1=54682$, $T_2$ increases to 57510 (left), or for fixed $T_2=57510$,
$T_1$ decreases to 54682 (right).
Events are selected using an $R<2^\circ$ aperture, and each event is weighted
by the probability. Three lines are for different energy bands:
200\,MeV--100\,GeV (black), 200\,MeV--1\,GeV (blue), and 1\,GeV--100\,GeV (red).
\label{fig:fig3}
}
\vspace{0mm}
\end{figure*}

	As expected, no on-pulse interval shows significant modulation.
However, the off-pulse light curves show modulation consistent with
that reported by \citet{xw15}, deserving further study.
Because we use a source-weighted photon analysis we did not need to use
the very small apertures of that study. This avoids the extreme sensitivity
of modulation significance to the aperture choice.
With a conservative 2$^\circ$ aperture, we compute the cumulative
weighted H-test significance \citep[][]{k11} starting from the beginning or end of
the data set; a significance which grows as more exposure is included
indicates a real, steady signal. We track
the modulation in low (200\,MeV--1\,GeV) and high (1\,GeV--100\,GeV) energy bands in addition
to the full LAT range. The cumulative significance curves are shown in Figure~\ref{fig:fig3},
and the final low- and high-band light curves are shown in Figure~\ref{fig:fig2}.
We further changed the aperture size (e.g., $R=3-5^\circ$)
and found that the results do not change significantly. Although larger counts for larger
apertures may result in higher significance especially at lower energies with the
probability weighting, the weighting may not be accurate at the event-by-event level for
timing analysis, hence an improvement in significance is not necessarily expected.
As noted above the exposure variation rapidly becomes negligible, dropping to the percent level
after $\sim$100\,days of exposure. Monte Carlo simulations confirm that including this
small variation in the bin exposure makes negligible difference to the H-test
significance.

	We note that the nearby flaring source 3FGL~J1316.0$-$3338 (hereafter J1316)
at a distance of $d\sim1^\circ$ may affect
the results. Although it is not very likely that this source exhibits periodic modulation
on J1311 periods, a short-time flare of J1316 may be a concern. This effect
is minimized because we use probability weighting, but there may be some residual effects.
We therefore produced a light curve for J1316\footnote{see also https://fermi.gsfc.nasa.gov/ssc/data/access/lat/FAVA\\/LightCurve.php}
and found two relatively large flares. We excised the time intervals for the these flares
($\pm7$\,days), and re-computed the
H-test significance for J1311 in the three energy bands as we did above.
We find that the results do not change whether or not we excise the J1316 flare intervals.

	The low-band signal fades at late times while
the high-band signal continues the secular increase in significance;
this suggests some unknown, slowly varying, soft contamination.
The full-band signal
shows the largest significance, reaching $p\lapp 2 \times 10^{-4}$. This significance
decreases slightly, with $\sim 4.5 \times 10^{-4}$ by the end of the data set;
this is as expected since the low-band counts dominate. This $\sim 10\times$
improvement in significance over the  \citet{xw15} result is encouraging, but
requires further exposure to be definitive. We can, of course, improve exclusion
of low-energy and pulse contamination, by varying phase, aperture, time and energy
cuts. As can be seen in Figure~\ref{fig:fig1}, using a narrower phase interval
may suppress contaminations of the pulse peaks better and hence reveal stronger modulation.
We therefore varied the selection criteria including the phase interval, and find the the
results are not very sensitive to the selection criteria.
For example, using $\phi_p=0.182-0.549$, aperture $R=1.27^\circ$,
$E_\gamma$=1.01--100\,GeV and interval MJD=56047--57431, we reach $p\sim10^{-6}$ significance.
Interestingly, this `tuned' high-energy light curve is nearly
100\% modulated. However, given that the post-trials significance is
$\gapp 10^{-4}$, comparable to that for the initial un-tuned cuts, we cannot claim that
such large modulation is required.

	For phase-resolved spectral analysis,
we generated the phase-resolved spectra for four different pulse-phase intervals:
off-pulse (0.127--0.587), P1 (0.597--0.737), interpulse (0.757--0.917), and P2 (0.937--1.107)
intervals, and performed likelihood analysis using {\tt pylikelihood} with the data
of each pulse interval. For these, normalizations for all the sources in the 3FGL
model file were scaled down by the phase interval and the resulting flux for J1311 was
scaled up by the same amount.
The results are rather insensitive to the precise
definition of these phase intervals. Here, we held all the parameters
(except for those of PLEXP for J1311) fixed at the best-fit values obtained
in the phase-averaged analysis above. The results are shown in Table~\ref{ta:ta1}.
We further tried to see if the result changes if we fit the
diffuse backgrounds ({\tt gll\_iem\_v06.fits} and {\tt iso\_P8R2\_SOURCE\_V6\_v06.txt}) as well,
and found that the results do not change significantly.

\begin{figure}
\centering
\hspace{-5.0 mm}
\includegraphics[width=3.3 in]{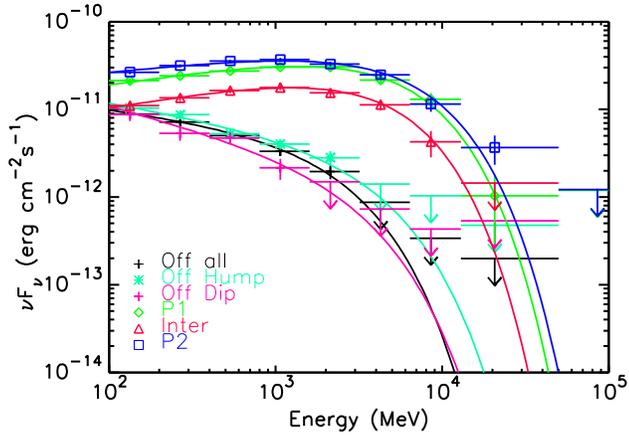}
\figcaption{{\it Fermi}-LAT SED of J1311. Flux in each energy band was
measured by fitting the amplitude of the best-fit model (see Section~\ref{sec:sec3_1})
in that band. When the $TS$ value for the fit is less than 5, the 95\% flux upper limit was derived
by scanning the amplitude until log$\mathcal L$ changes by 1.35 with the {\tt UpperLimits.py} script
provided along with the {\it Fermi}-LAT Science Tools.
\label{fig:fig4}
}
\vspace{0mm}
\end{figure}

	For the off-pulse interval, we fit both the total flux and the spectra from the
``hump'' phase ($\phi_B=0.6-1.2$) and the lower ``dip'' phase ($\phi_B=0.2-0.6$).
The spectral parameters are not significantly different although
the hump power law is nominally slightly harder (Table~\ref{ta:ta1}).
Note that the cutoff energy $E_{\rm c}$ for the ``dip'' phase is not well constrained, so we
hold $E_{\rm c}$ at the value obtained for the ``hump'' phase.
The spectra are shown in  Figure~\ref{fig:fig4} and the fit parameters are presented in Table~\ref{ta:ta1}.
We have also fit a simple power-law (PL) model to the off-pulse phase spectra.
Using the log-likelihood fit statistic,
we find that PLEXP fits better than PL does
with $\Delta \mathrm log\mathcal L$ of 9, 1, and 7 for ``all'', ``dip'', and ``hump'',
respectively, in the 100\,MeV--300\,GeV band.
For the  ``all'' and the ``hump'' spectra, a curvature (i.e., a cutoff) is still required ($\gapp$3$\sigma$),
but we cannot tell whether the ``dip'' spectrum requires a curvature.
Nevertheless, the shapes are very different from the on-pulse spectra (Fig.~\ref{fig:fig4}).

          Because the low energy light curves' significance appears concentrated in the early data,
we also investigate long-term source variability.  The variability index \citep{fermi3fgl} in the
3FGL catalog is 52, and hence the total source flux has no significant variability. Variability
might still be significant in some phase bins. As in section~\ref{sec:sec3_1}, we selected
4 phase intervals and constructed a light curve for each phase interval, with
time bin 2\,Ms for the faint off-pulse interval and 1\,Ms for the rest.
Note that there is a nearby variable source (J1316)
which can contaminate the J1311 light curves. For each time bin, we performed likelihood
analysis using {\tt pylikelihood} to fit the amplitudes of J1311 and J1316.
All other source and background parameters are held fixed at those obtained by
a pulse-phase-resolved analysis (section~\ref{sec:sec3_1}). We used the best-fit source fluxes to
calculate $\chi^2$ for a constant flux. As expected the on-pulse phases dominated by magnetospheric
emission are consistent with steady flux. The off-pulse phase gives $\chi^2/{\rm dof} = 57/86$, so even
this interval is consistent with steady emission. We conclude that there is no
long-term time variability in J1311, which is consistent with the study of \citet{tjlp+17}.

\subsection{X-ray variability and Spectrum}
\label{sec:sec3_2}
        J1311 is known to exhibit optical and X-ray flares \citep{r12,kyk+12}, and a
possible orbital modulation in the X-ray band has been reported \citep[][]{kyk+12,apg15}.
So a re-examination of the X-ray observations provides an important comparison with the gamma-ray modulation.

\begin{figure*}
\centering
\begin{tabular}{cc}
\hspace{-7.0 mm}
\includegraphics[width=3.35 in]{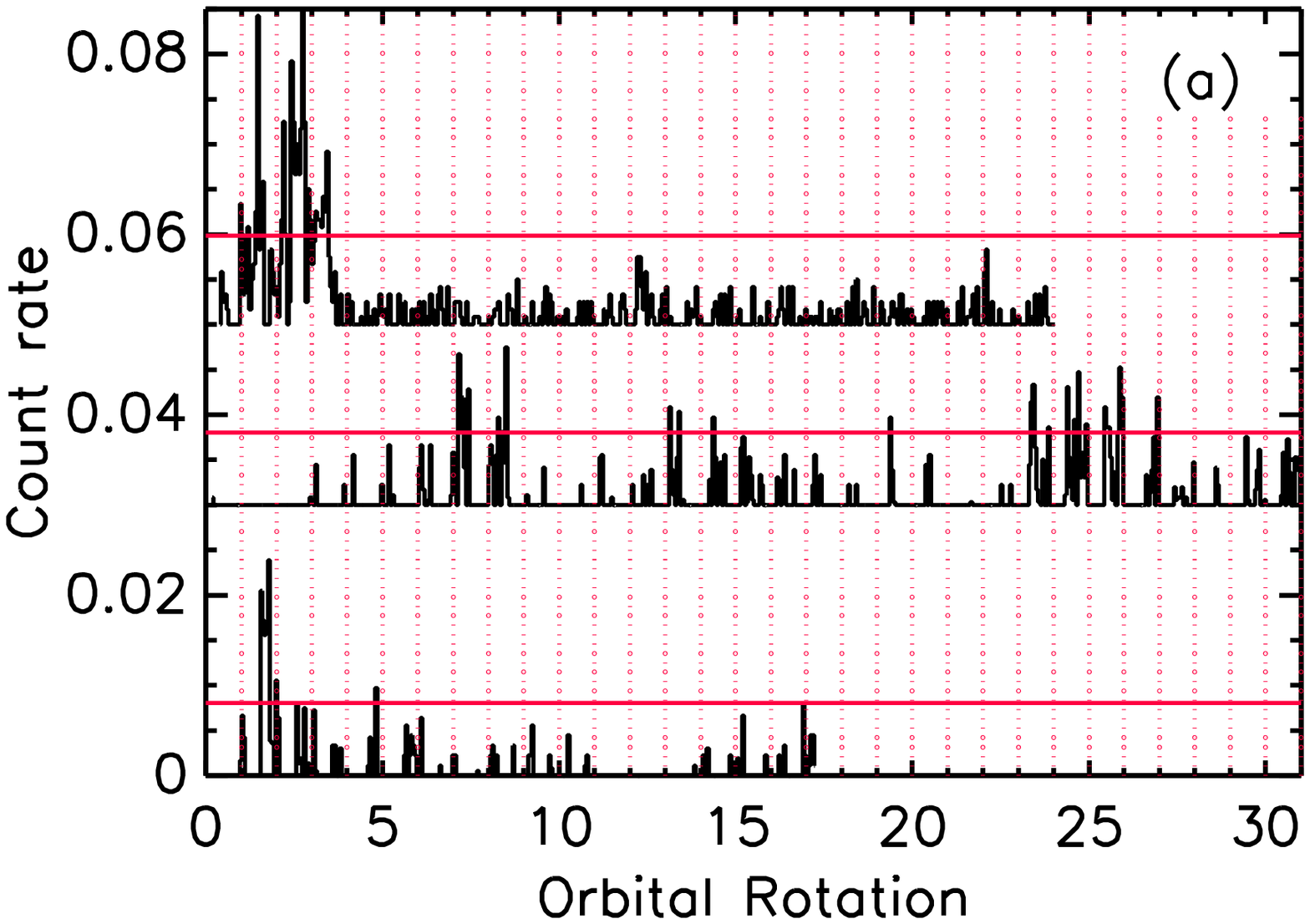} &
\includegraphics[width=3.6 in]{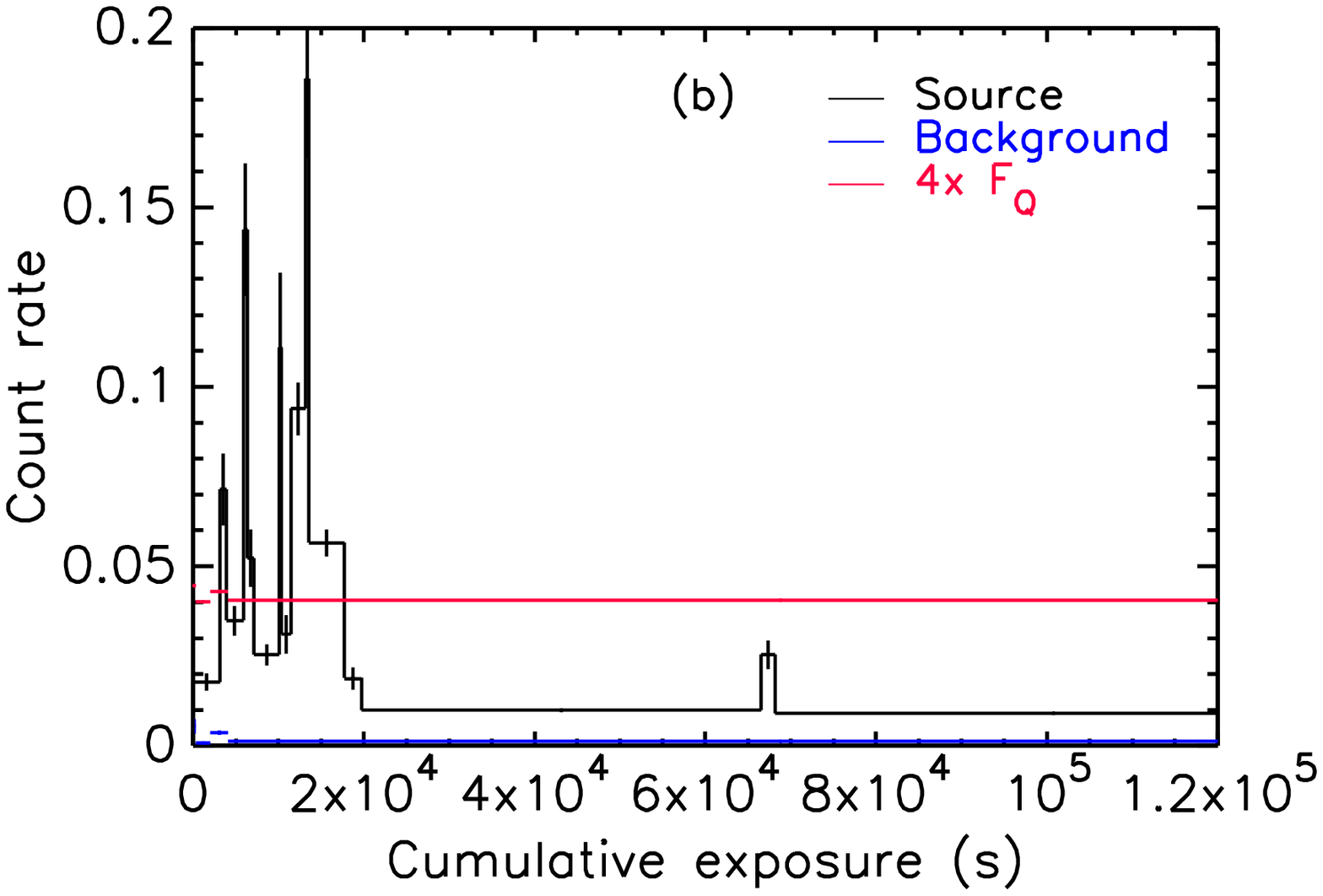} \\
\hspace{-2 mm}
\includegraphics[width=3.5 in]{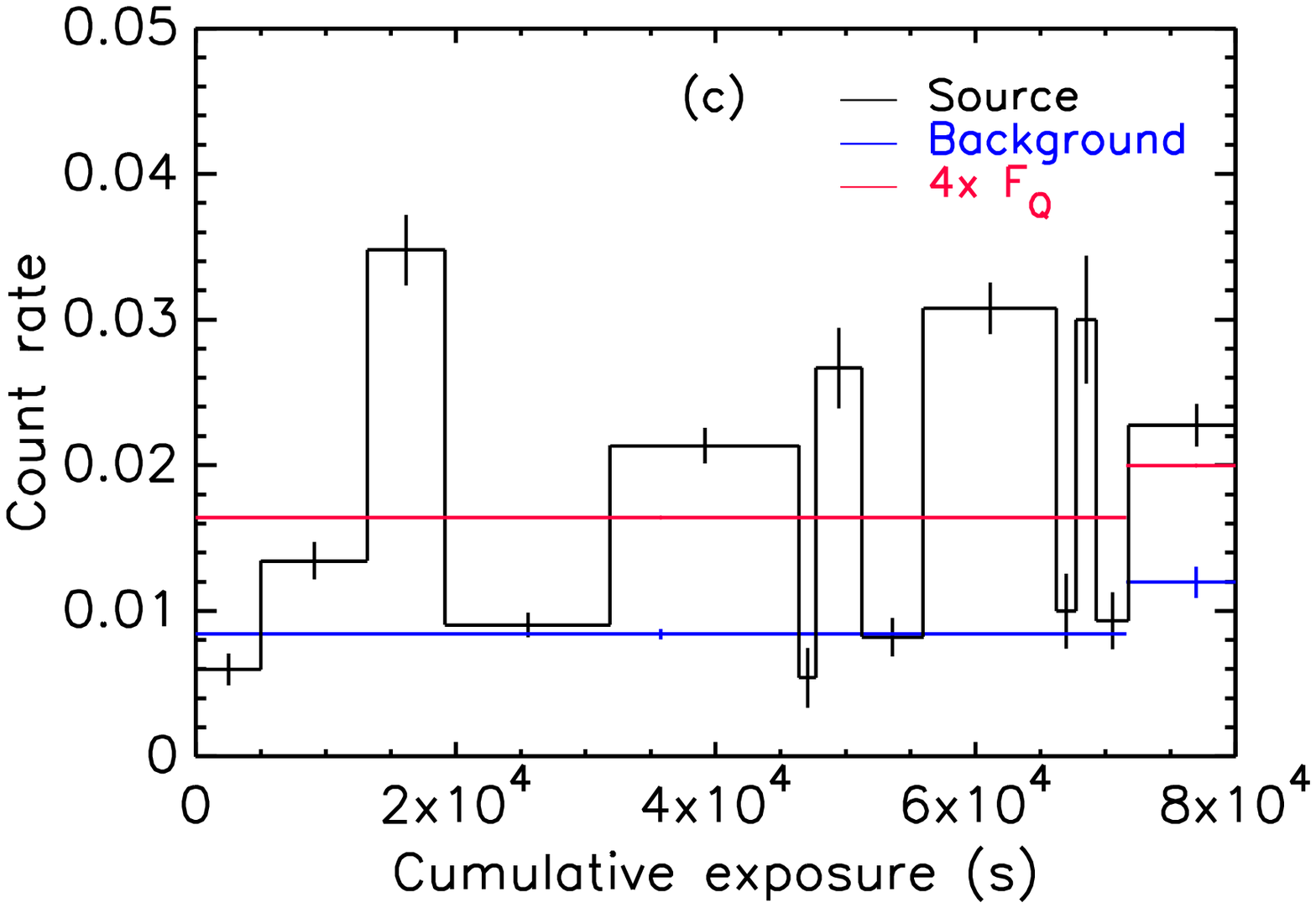} &
\includegraphics[width=3.5 in]{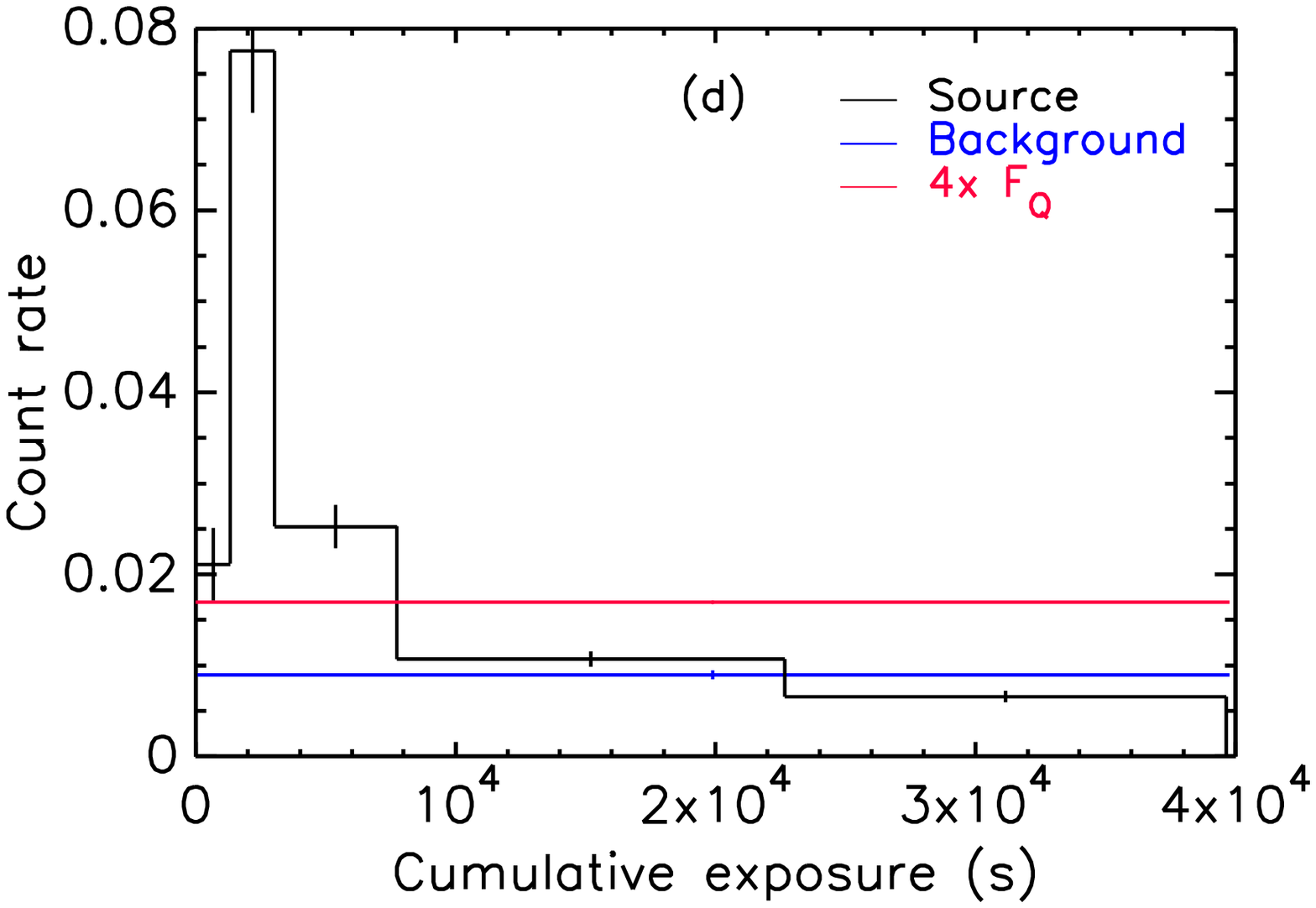} \\
\hspace{-7 mm}
\includegraphics[width=3.3 in]{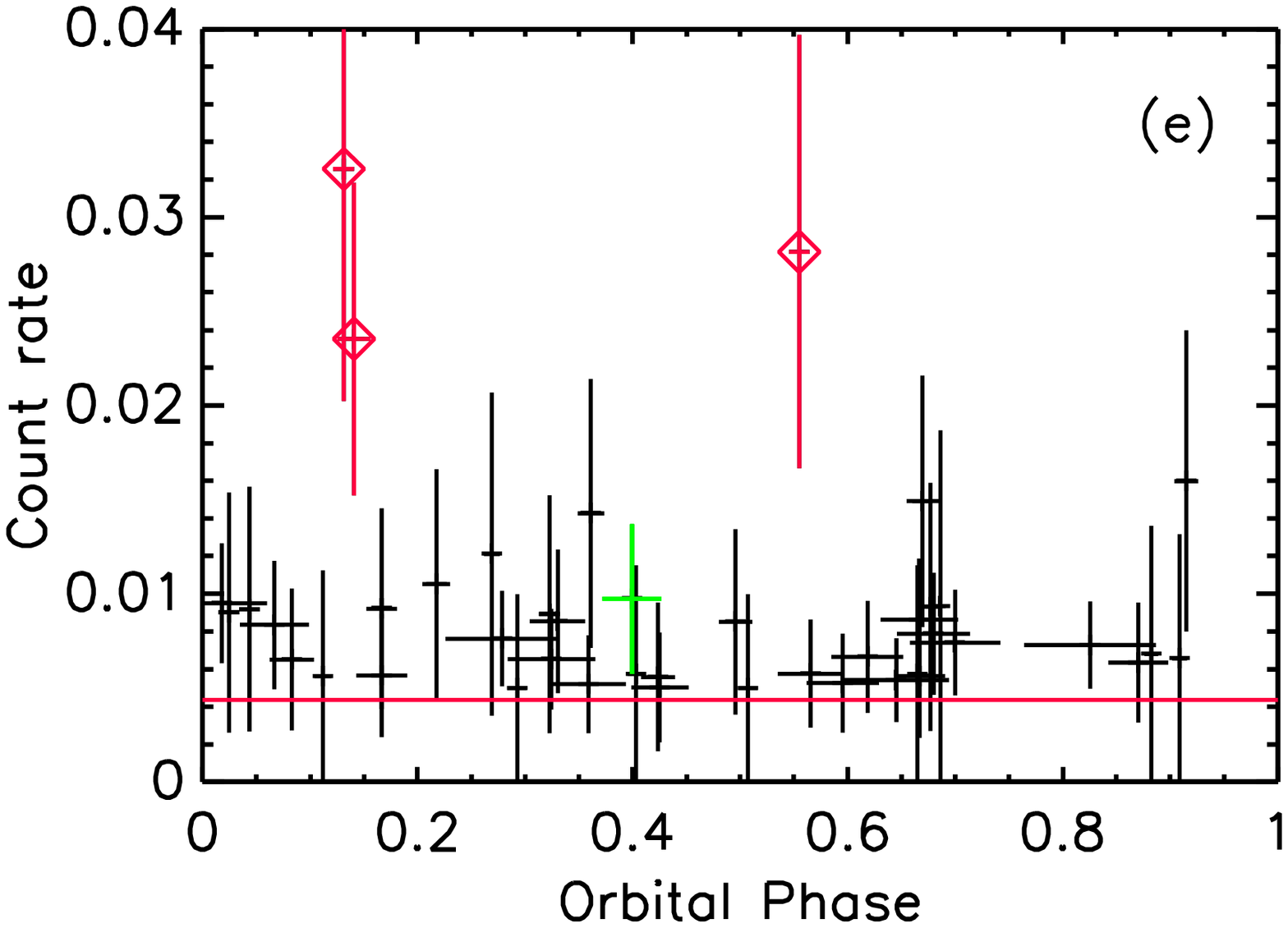} &
\hspace{-5 mm}
\includegraphics[width=3.45 in]{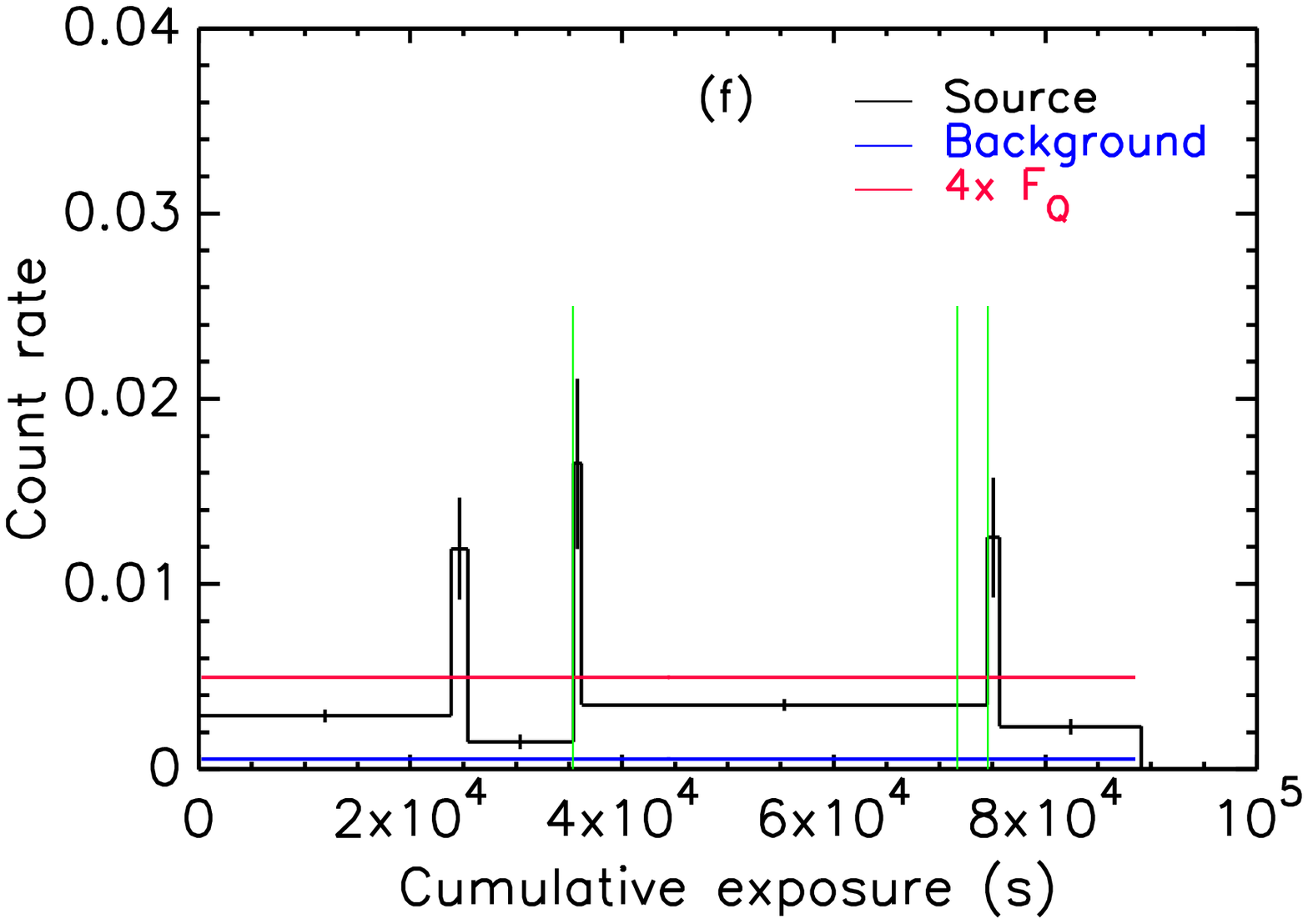} \\
\end{tabular}
\figcaption{(a): Unfolded 0.3--10\,keV X-ray light curves of {\it XMM-Newton}/MOS,
and {\it Suzaku}/XIS observations (300-s time bins, background subtracted).
{\it XMM-Newton}, {\it Suzaku} AO6 and AO4 from the top. Data are offset,
and amplitudes are scaled down by a factor of 4 ({\it XMM-Newton}) or 3 ({\it Suzaku}) for clarity.
Red vertical lines mark phase 0.
Note that the 96-minute {\it Suzaku} orbit is visible as gaps in the plot.
(b)--(d): Bayesian-block representation of the light curves for the {\it XMM-Newton} (b),
{\it Suzaku}/AO6 (c) and {\it Suzaku}/AO4 (d) data. Data gaps present in
the data are removed in this representation to avoid artifacts produced due to
the gaps \citep[see][]{snjc13}. Backgrounds are shown in blue, and $4\times$ the quiescent
level is shown in red.
(e): A folded {\it Swift} light curve made of data points in which the flux is
greater than $4\times$ the quiescent level. We denote
data points which are significantly higher (with 90\% confidence) than
4$\times$ the quiescent level in red. The green point corresponds to the first high-flux
point in panel f.
(f): Bayesian-block representation of the unfolded {\it Swift} light curve with the data
gaps removed. Vertical green lines correspond to the high-flux points (red) in (e).
\label{fig:fig5}}
\vspace{0mm}
\end{figure*}

        We re-examined archival X-ray observations taken with {\it Swift}, {\it XMM-Newton} and
{\it Suzaku}. Source fluxes were extracted from apertures with
radius $R=40''$, $R=16''$, and $R=40''$, respectively, while the background was monitored
by nearby source-free apertures of radius $R=60''$, $R=32''$, and $R=60''$.
Event times were barycentered with the ephemeris of  section~\ref{sec:sec3_1}.  In Figure~\ref{fig:fig5}
we plot the binned ($\Delta t=300$\,s) {\it XMM-Newton} and {\it Suzaku} flux time series, with $t=0$ set
at the ascending node prior to observation start. We use {\it Chandra} observations for
spectral analysis only because of the short exposures and small number of counts. The source and
background events are extracted using a $R=3''$ aperture and an annular region with $R_{\rm in}=5''$
and $R_{\rm out}=10''$, respectively.

	We investigate source variability in the time series
using the Bayesian block algorithm \citep[][]{s98,snjc13}
implemented in the {\tt python} {\tt astroML} package \citep[][]{astroML}.
The algorithm computes the number of optimal blocks and the change points for the time series.
In the source time series, we find 15, 13, 5, 7, and 2 blocks for
{\it XMM-Newton}, {\it Suzaku}/AO6, {\it Suzaku}/AO4, {\it Swift}, and {\it Chandra};
there are some blocks with much larger count rates than the minimum level and others
similar to the minimum level (Figure~\ref{fig:fig5}).
We performed the same analysis with the background time series to see if some of the source
variabilities can be attributed to background activities.
The {\it Suzaku}/AO4, {\it Swift}, {\it Chandra} backgrounds are well explained
with a single block (i.e., no variability),
and the others require 2--4 blocks suggesting some variability in the background; these
variabilities are small and do not seem to correlate with the source activity.
Furthermore, the background flux is only a small fraction of the source flux, hence the small background
variabilities are unlikely to drive the source activities.

       Strong flaring variability is clearly visible (Figure~\ref{fig:fig5}), with episodes reaching
$>10\times$ the quiescent flux. We have marked some flux levels
to help guide the eye at 4$\times$ the quiescent flux (red horizontal lines).
There is no obvious preferred phase for flaring events. It appears, especially
in the {\it XMM-Newton} data, that the flares can be clustered in episodes lasting
several orbits $\sim 10-20$\,ks.

	The {\it Swift} data are snapshots over many years, so all we can show
is the phase dependence of intervals (100--1400\,s) where the flux was
$>4\times$ the quiescence (Figure~\ref{fig:fig5} e).
In Figure~\ref{fig:fig5} e, for only a few intervals was this excess more than 90\% significant
(red points in Figure~\ref{fig:fig5} e). For the significance, we calculate
the Poisson probabilities of having the observed counts in the time bins
greater than the background plus $4\times$ the ({\it XMM}-estimated) quiescent counts,
considering the trial factor (number of data bins in the figure).
We denote these time bins in green lines in the Bayesian-block
light curve as well (Figure~\ref{fig:fig5} f). Two of the high-significance data points
in Figure~\ref{fig:fig5} e have a corresponding peak in Figure~\ref{fig:fig5} f.
The lowest one does not have a counterpart (green line at $T\approx7\times10^{4}$\,s
in Figure~\ref{fig:fig5} f); this block has slightly elevated flux
compared to the minimum, suggesting low level variability. In Figure~\ref{fig:fig5} f,
the first flare does not have a high-significance (red)
counterpart in Figure~\ref{fig:fig5} e; the corresponding point is denoted as a green cross in
Figure~\ref{fig:fig5} e. Again there is no clear phase grouping of the limited number
of significant flares.

       {\it XMM-Newton} has the highest count rate and the longest continued coverage.
To test for orbital modulation we formed the light curve for ``flare'' ($T<56872.1$ MJD,
the first $\sim$4 orbits in Figure~\ref{fig:fig6} left)
and ``quiescent'' ($T>56872.1$ MJD) periods (Figure~\ref{fig:fig6} left)
The quiescence light curve is fully
consistent with constant flux ($p=0.5$); no variability
is seen in our Bayesian-block analysis.
With limited flare events in
{\it XMM} alone the light curve is spiky, but we cannot discern a preferred phase.
Substantially longer integrations are needed to infer the detailed flare
behavior -- but the data already show that flares can start at any phase.

       The  {\it XMM-Newton} data provide by far the best statistics for measuring the
X-ray spectrum. We can distinguish between the ``flare'' epoch and ``quiescent'' epochs.
For each, we extracted source and background events using
apertures of $R=16''$ and $R=32''$, respectively. We then calculated the response files
using the standard tools of SAS version 20141104\_1833 ({\tt rmfgen} and {\tt arfgen}).
We then jointly fit the two 0.3--10\,keV spectra with an absorbed power-law model having
a common absorbing column density ($N_{\rm H}$). The source absorption is undetectably small;
it is consistent with 0 with the current statistics.
Unsurprisingly, the flare spectrum is substantially harder than that in quiescence
(Table ~\ref{ta:ta3}).

       While other X-ray data sets provide insufficient statistics for accurate fits
we can at least estimate fluxes and hardness. In all cases, we fix $N_{\rm H}= 1 \times 10^{20}\rm cm^{-2}$.
For the {\it Suzaku} data we use $R=40''$ and $R=60''$ for the source and background apertures,
respectively, and response files generated with {\tt XSELECT}. The AO4 data had
a clear bright flare and the separate flare and quiescence spectra have indices similar
to the {\it XMM} measured values.  
In AO6 the source is relatively hard and bright when compared to the quiescent state
observed with {\it XMM-Newton}, suggesting substantial flare contributions
(Table~\ref{ta:ta3}; see also Figure~\ref{fig:fig5} c).
In the first of two archival {\it CXO} ACIS
data sets the source is relatively bright with substantial flare contribution; the second
is closer to quiescence. The spectral index uncertainties are too large to select a state.
For {\it Swift} we simply combined all XRT observations, used pre-processed files for
the source spectra and separately created background spectra using $R=60''$
apertures in the source-free regions. A fit to the merged spectrum gave an intermediate
level average flux and a poorly determined, but hard $\Gamma$, again suggesting substantial
contribution from poorly measured flares. Note that if the {\it XMM} flare/quiescence
flux ratio is typical, a flare duty cycle as small as 10\% will cause significant
contamination in the spectral fits.

\newcommand{\marka}{\tablenotemark{a}}
\newcommand{\markb}{\tablenotemark{b}}
\begin{table}[t]
\vspace{-0.0in}
\begin{center}
\caption{Absorbed Power Law Fits to the X-ray data}
\label{ta:ta3}
\vspace{-0.05in}
\scriptsize{
\begin{tabular}{cccc} \hline\hline
State         & $N_{\rm H}$      &   $\Gamma$     &  $F_{\rm 0.3-10\,keV}$ \\
              &   $10^{20}\rm cm^{-2}$ &   $\cdots$     &  $10^{-14}\rm erg\ cm^{-2}\ s^{-1}$ \\ \hline
{\it XMM} flare         &   $1\pm2\marka$              &  $1.26\pm0.07$ & $47\pm2$ \\
{\it XMM} quiescence    &   $1\pm2\marka$              &  $1.67\pm0.09$ & $6.6\pm0.3$ \\ \hline
{\it Suzaku} AO4 &   $1\markb$  &  $1.3\pm0.1$ & $26\pm2$ \\
{\it Suzaku} AO4 Flare &   $1\markb$  &  $1.19\pm0.15$ & $94\pm10$ \\
{\it Suzaku} AO4 Quie. &   $1\markb$  &  $1.6\pm0.2$ & $14\pm2$ \\
{\it Suzaku} AO6 &   $1\markb$  &  $1.16\pm0.05$ & $40\pm2$ \\ \hline
{\it Chandra} 11790 &   $1\markb$  &  $1.3\pm0.3$ & $18\pm4$  \\ 
{\it Chandra} 13285 &   $1\markb$  &  $1.4\pm0.3$ & $8\pm2$ \\ \hline
{\it Swift} combined &   $1\markb$  &  $1.20\pm0.15$ & $19\pm2$ \\ \hline
\end{tabular}}
\end{center}
\vspace{-0.5 mm}
$^{\rm a}${Tied.}\\
$^{\rm b}${Frozen.}\\
\end{table}

\subsection{Optical variability}
\label{sec:sec3_3}

        The optical light curve has the dramatic $\sim 4$ magnitude ($\sim 40\times$)
orbital modulation characteristic of BW pulsars, with maximum at
pulsar inferior conjunction $\phi_B=0.75$ when the heated face of the companion is
best visible. In addition it shows optical flares with rise times as short as
$\sim 300$\,s and amplitudes as large as $\sim 6 \times $ the peak flux \citep{r12}.
We would like to know how these optical events relate to the X-ray flares above.

        We are fortunate in having simultaneous optical data from the {\it Swift}/UVOT
and {\it XMM-Newton}/Optical monitor (OM). Unfortunately, these data were taken with
a variety of integration times and filters. The {\it XMM-Newton}/OM exposures were 2\,ks
($\Delta \phi_B=0.36$). This coarse time sampling makes it difficult to identify any
but the largest flares.  The {\it Swift}/UVOT exposure varied between 30\,s and 1.6\,ks.
To extract fluxes for an initial study of the optical behavior we used standard
tools ({\tt uvotsource} and {\tt omichain}; Section~\ref{sec:sec2}).
For the {\it Swift} data, we barycenter correct the exposure midpoint times.
For the {\it XMM-Newton} data, we used the
barycenter-corrected X-ray photon arrival times, and interpolate them to obtain
the barycenter-corrected OM times. The {\tt omichain} automatically detects the source
in the {\it XMM-Newton}/OM data and properly calculates the source flux.
For the {\it Swift}/UVOT data, we used a $R=5''$ and a $R=15''$ apertures for the
source and the background, respectively, to measure the source flux.

\begin{figure*}
\centering
\hspace{-7.0 mm}
\begin{tabular}{cc}
\hspace{2.0 mm}
\includegraphics[width=3.36 in]{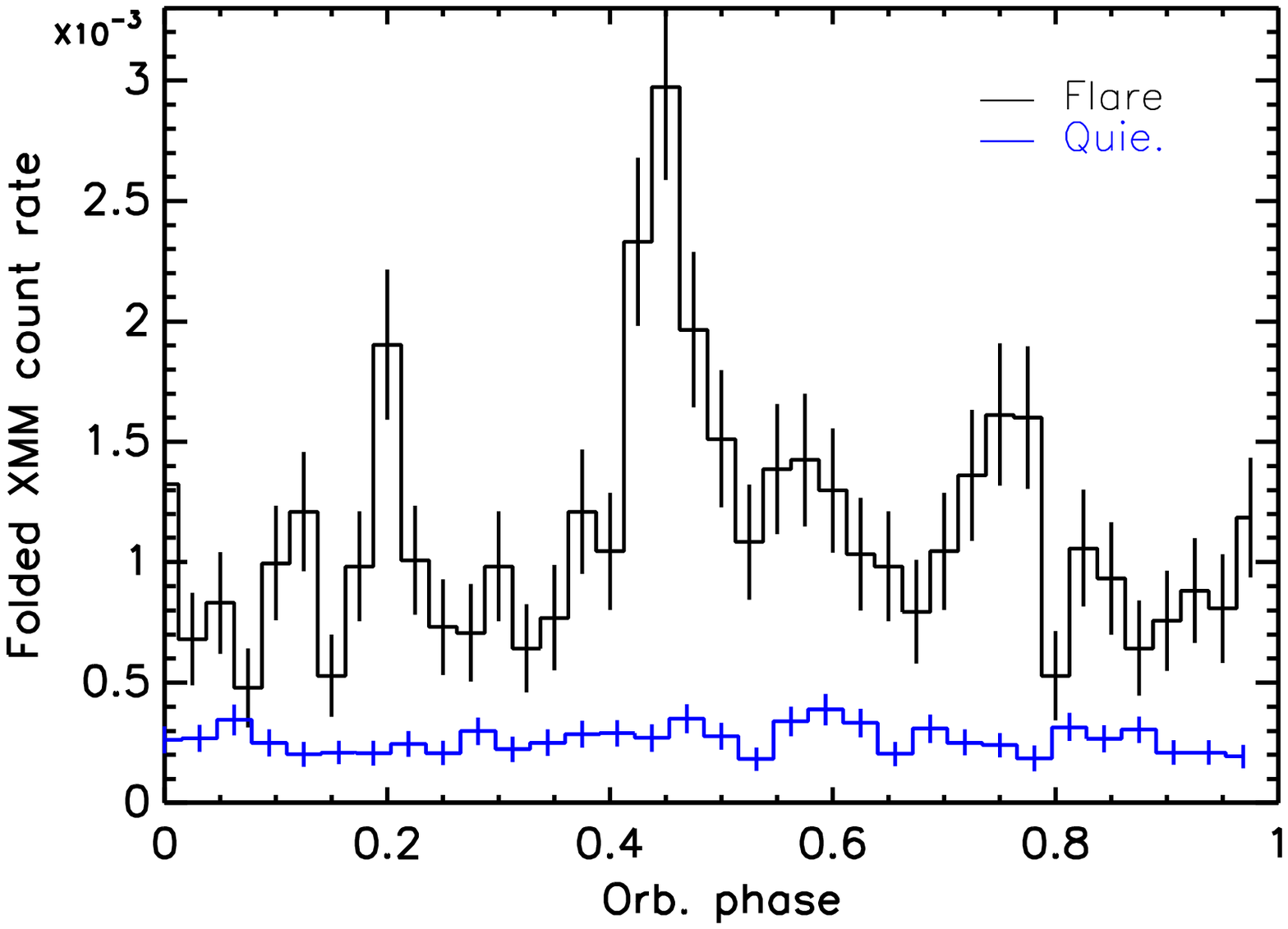} &
\includegraphics[width=3.4 in]{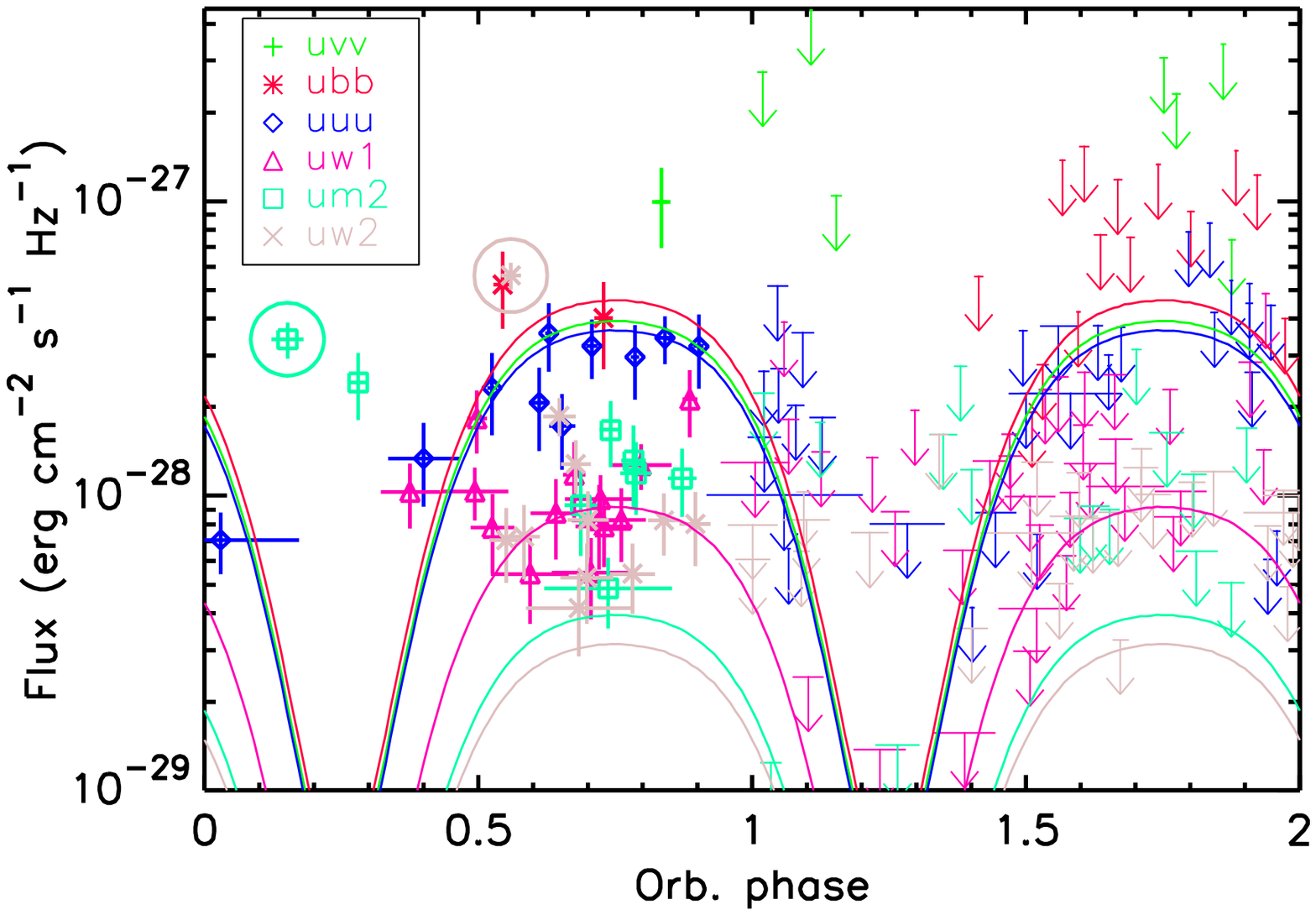}\\
\end{tabular}
\figcaption{Left: Folded X-ray orbital light curve measured with {\it XMM-Newton}/MOS
during quiescence (blue, 113\,ks) and flare (black, 20\,ks).
There are more events in the quiescent light curve due to the longer exposure.
Right: Folded optical-band light curves measured with {\it Swift}/UVOT in
the $V$, $B$, $U$, $UVW1$, $UVM2$, and $UVW2$ bands. Data points with error bars are for
exposures in which J1311 is detected (first period). When the source is not
detected, flux upper limit is shown (second cycle). Data points corresponding
to the XRT high-flux states (red in Figure~\ref{fig:fig5} e) are denoted with a circle.
\label{fig:fig6}
}
\vspace{0mm}
\end{figure*}

        The resulting barycentered midpoints of the {\it Swift} observations are
phased with the binary ephemeris and plotted in Figure~\ref{fig:fig6}, with detections, mostly near $\phi_B=0.75$,
in the first cycle and upper limits in the second cycle. To know if these detections
are bright compared with the quiescent state we need to compare with the quiescent
(pulsar heating) light curve. We can use the ground-based observations of \citet{rfb15} to
interpolate quiescent $V$ and $B$ light curves. However the quiescent UV fluxes of J1311 are
not known. The hot $T_{eff}\approx 12,000$\,K face of the He companion must contribute
some flux. However, after de-reddening by an estimated $E(B-V)=0.173$
\citep[Galactic average towards J1311;][]{rfb15},
we find that the phase average optical UV spectrum does not match a simple
blackbody. This is in contradiction to \citet[][]{kyk+12}'s UVOT analysis.

         Still, we need an estimate of the quiescent flux in the UV filters and so, to
be conservative, we scaled the best-fit model $B$ light curve assuming a simple blackbody
spectrum across the UVOT bands to generate
$U$, $UVW1$, $UVM2$, and $UVW2$ light curves; these estimates are
shown in Figure~\ref{fig:fig6} and
also consistent with the {\it Swift} upper limits.  Note that the modest  {\it Swift}
$V$ sensitivity means that the source was only detected in a high (flare) state. One
$B$ detection also likely represents a mild flare. In the $UVM2$ and $UVW2$ bands, the
measured fluxes are in general higher than the blackbody-extrapolated model fluxes. This
may suggest that the source spectrum at high frequencies deviates from a simple blackbody.
Several UV detections during the
`night' phase ($\phi_B=0.0-0.5$) clearly represent a large increase over the expected
quiescent flux. The {\it XMM-Newton}/OM data are very coarse and have few detections,
yet we can identify several bright flare points here, as well.

\subsection{Variability Statistics and the Optical/X-ray Correlation}
\label{sec:sec3_4}

\begin{figure}
\centering
\hspace{-7.0 mm}
\includegraphics[width=3.3 in]{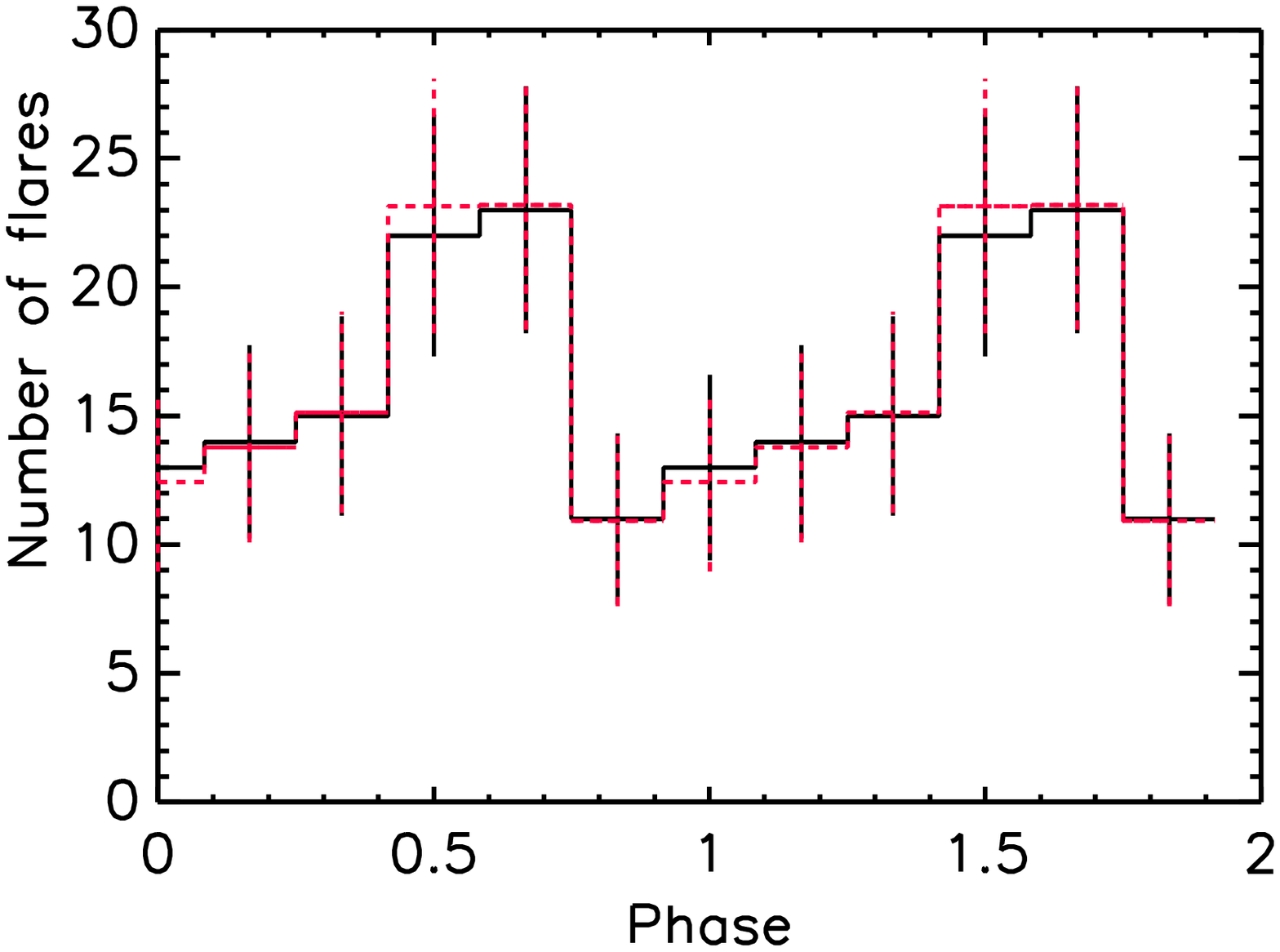} \\
\hspace{-7.0 mm}
\includegraphics[width=3.4 in]{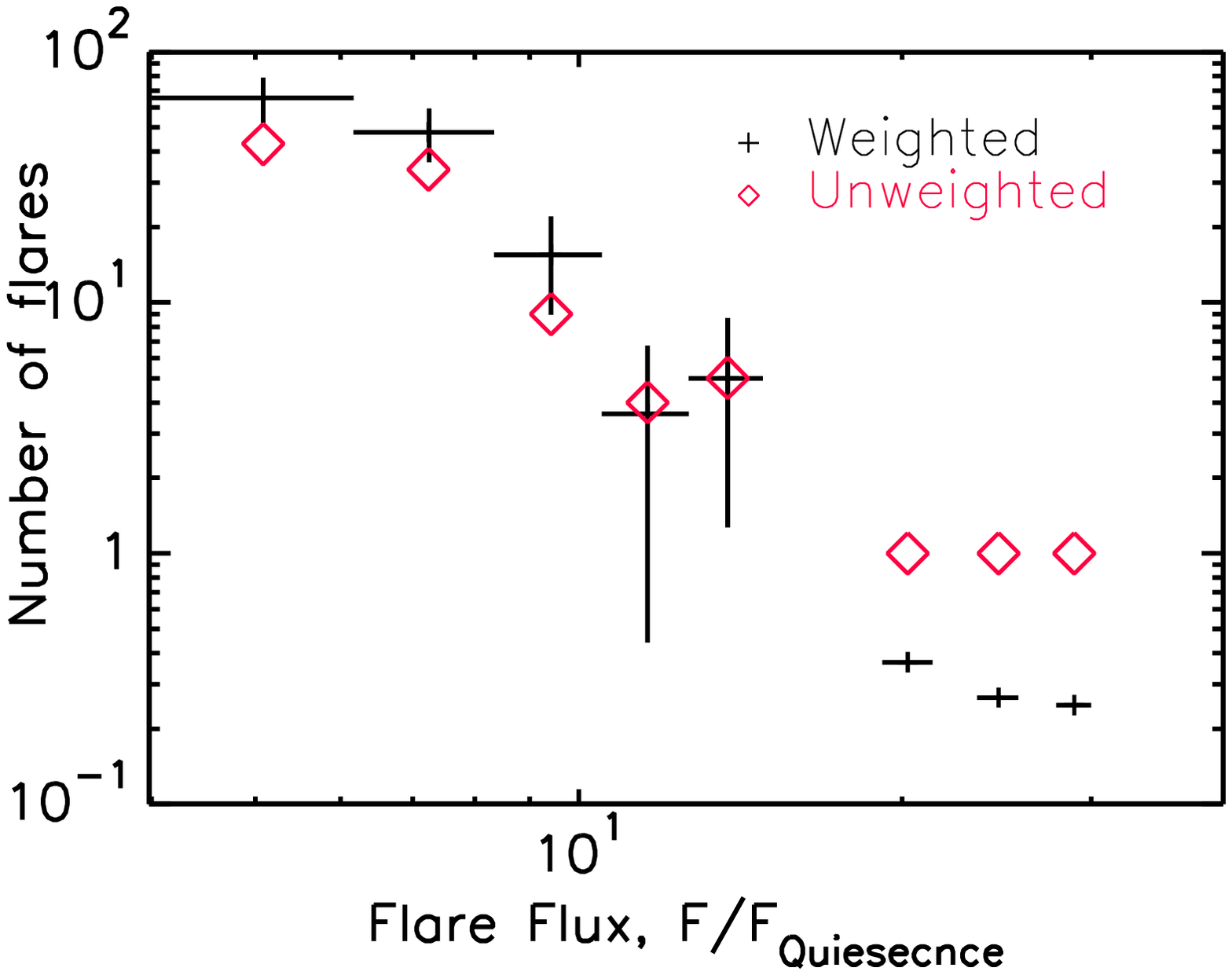}
\figcaption{{\it Top}: Number of flares detected in each phase in the X-ray observations
({\it XMM-Newton}, {\it Suzaku}, {\it Swift} combined). Red dashed line shows
exposure-weighted flare counts. {\it Bottom}: Flux distribution
of the detected flares. Black data points are made by weighting red points
with the flux uncertainties.
\label{fig:fig7}
}
\vspace{0mm}
\end{figure}

       We have identified flare intervals in each X-ray data set ($\Delta t\sim$300--400\,s)
for which the flux is $\gapp 4\times$ the quiescent value; adjacent flare bins are
counted as a single flare.
When finding the number of flares, we use the light curves made with constant time bins
instead of the Bayesian-block representation because we are concerned with the gap removal;
multiple flares separated by data gaps may be seen as one when the
gap is removed.
Note again that ``flare'' means the time intervals with measured flux greater than
$4\times$ the quiescent level in our definition.
We note that this definition for flare is arbitrary;
we use this 4$\times$ the quiescent level only for reference flux (e.g., for ``large'' flux), and hence
the ``flare'' definition should be considered with care.
We then histogram the start phase of these flare events
as a function of orbital phase (Figure~\ref{fig:fig7}).
While it is clear that flares can be visible at any phase, with this coarse binning
we see a mild, $\sim 50$\% enhancement at and just before $\phi_B= 0.75$, pulsar inferior
conjunction when we view the heated face most directly. However, this excess has low
($p=$0.2) significance, and so additional monitoring will be needed to determine the
flare phase distribution.
Note that the difference in exposure time for each phase bin in Figure~\ref{fig:fig7}
is $\sim$10\% (minimum to maximum), and
has no significant impact. We show the exposure corrected flare counts in Figure~\ref{fig:fig7}.

        We would like to know the duty cycle of this flaring activity.
For this, we use the Bayesian block representation of the light curves
(Figure~\ref{fig:fig5} b--d and f). In the light curves, we calculate
the count rates and the exposure times for each time bin.
If we adopt the $>4\times$ flare definition then fully 19\% of all intervals are
in a flaring state. However, many of these must be statistically insignificant fluctuations.
If we additionally require a 90\% significance for the interval to have a flux excess
using the Poisson statistic with the background and the {\it XMM}-estimated quiescent source flux,
the flare duty cycle drops to 7\%. Note that the duty cycle inferred this way is similar
to what we infer using the binned light curves.
We conclude that the true duty cycle at this $4\times$
threshold lies between these generous (19\%) and conservative (7\%) rates.

We also show the number of flares as a function of flux.
In Figure~\ref{fig:fig7} bottom, we plot the distribution of flare fluxes from all
observations, with the flare flux in units of of the quiescent flux for the given 
observatory. Of course some excesses are low significance, and to account for this
we also plot the flare flux weighted by the inverse of the flux uncertainties. This
better approximates the true distribution, which is evidently a rather steep
($N \sim fl^{-3}$) power law.

       Further, we can investigate the association of X-ray and optical flares of J1311.
Because {\it XMM-Newton} and {\it Swift} provide simultaneous optical observations,
data taken with these telescopes are particularly useful for this study.
The OM data of {\it XMM-Newton} were taken only with the $V$ filter during the
flare period, and we find that J1311 was detected only when the X-ray flaring activity is
strongest. The OM time bins are of course large, but during the early bright flare $V$-band
detections seem to span all phases, including `night' phases. We thus infer that these
$V$ detections are flare associated. The other OM filters were used during quiescence. The
detections in these bands are only during the day phase, suggesting that we see only the heating
modulation of the companion.

\begin{figure}
\centering
\hspace{-7.0 mm}
\includegraphics[width=3.4 in]{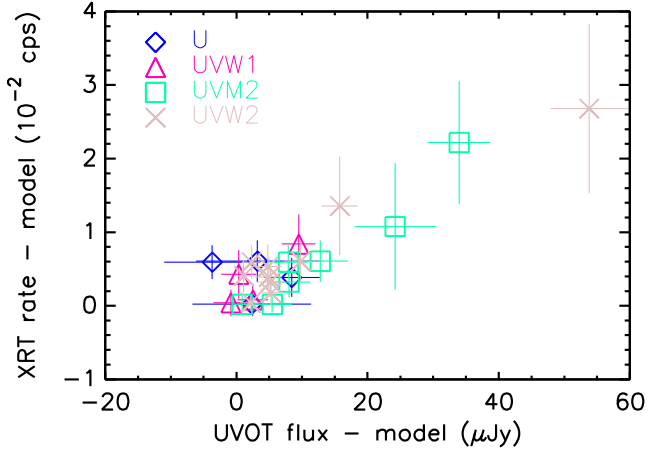}
\figcaption{UVOT flux vs. XRT count rates measured with {\it Swift}.
\label{fig:fig8}
}
\vspace{0mm}
\end{figure}

        The many {\it Swift} observations allow us to plot the excess above background X-ray flux
against the excess optical/UV flux beyond the heating model (Figure ~\ref{fig:fig8}).
The figure shows correlation between X-ray and optical fluxes. The significance
of the linear correlation is 7$\sigma$ (although this is, of course, sensitive to the
UV flux model that defines the flare excess).
If we separately consider each band,
the significance is highest in the $UVM2$ and the $UVW2$ bands, being $4\sigma$
and 5$\sigma$, respectively. We believe this suffices to show that the X-ray and optical/UV
flares are due to the same events. Recently \citet{hbt17} have noticed
correlated X-ray and optical variability in BW candidate XMMU~J083850.38$-$282756.8,
which may represent similar flaring events.

	Finally, we would like to know if these X-ray/optical flare events are associated
with the off-pulse GeV modulation shown in Figure~\ref{fig:fig2}. The low-significance
modulation in Figure ~\ref{fig:fig7} peaks at the rising part of the off-pulse LAT light curve,
so such association
is at least plausible. However, we are unable to associate individual flare events with
gamma-ray off-pulse activity. For $R<1^\circ$ from the source position,
energy $>$1\,GeV and pulse phase intervals of 0.127--0.587 we get a total count rate
0.17 day$^{-1}$, including background. However, the integrated flare
time from all X-ray observations to date is 1.1\,day. None of these gamma rays arrives during
an X-ray flare, but the Poisson probability of this is 80\%. In fact, given the
substantial duty cycle for the flares, even if all off-pulse LAT photons arose during
flare periods, we would expect $\sim 1$ coincident count. We attempted to boost the
off-pulse count rate, by opening the energy range to $\gapp$100\,MeV (1.4 off-pulse
photons/day), but this gives only 1.6 expected coincident counts (with a substantial
fraction being background events). Two events are actually seen, no strong excess.
Our only inference is that the {\it Fermi}-LAT flux during the X-ray flare intervals is
$\lapp$10$\times$ the quiescent flux (90\% upper limit).
Much more extensive (30--100d) X-ray or optical flare
monitoring is needed to test for an event-level gamma-ray association. Thus we cannot
make a direct connection and, for the foreseeable future, only an indirect connection
of associated phase seems possible.

\section{Discussion and Conclusions}
\label{sec:sec4}
         We have analyzed optical to gamma-ray observations of J1311.
As suggested by \citet{xw15}, orbital modulation seems to exist above 200\,MeV in the off-pulse
interval. Our analysis shows that the low-energy modulation ($\lapp$1\,GeV) weakens at later times
($p\approx10^{-3}$ at 1600\,days to $2\times 10^{-2}$ at 3000\,days; the blue line in Figure~\ref{fig:fig3} left)
while high-energy ($>$1\,GeV) modulation persists. {This is puzzling because
relative to the constant pulsar emission (Figure~\ref{fig:fig4}) the off-pulse emission is
stronger at lower energies, hence larger modulation at lower energies is expected,
which we do not see. However, if the modulation is driven by the bulk plasma motion in the IBS,
stronger modulation in the high-energy band is possible because high-energy emission
is produced only in limited phase intervals \citep[e.g.,][]{ar17}.
Nevertheless, more data are required to establish firmly the gamma-ray modulation in J1311.

The overall off-pulse PLEXP spectral index
$\Gamma_{\rm 1}=2.24$ is steeper than values seen for LAT MSPs \citep[][]{fermi2pc};
thus is unlikely to be magnetospheric
in origin, suggesting perhaps variable shock (likely IBS) emission.

        The unpulsed source flux is maximum at the orbital phase $\sim$0.8 and minimum
at $\sim$0.4 in the gamma-ray band (Figure~\ref{fig:fig2} c). In massive star IBS (e.g.
gamma-ray binaries) the observed gamma-ray modulation is believed to be primarily due to
varying wind density (and hence shock stand-off, density and photon illumination) around 
an elliptical orbit.
However, J1311 like other BW-type binaries has negligible eccentricity, and so
the light curve peaks cannot be ascribed to orbital variation of the stars' separation.
Instead we must appeal to anisotropic emission and/or Doppler boosting in the relativistic
post-shock flow.

       In IBS emission models \citep[e.g.,][]{ar17,whvb+17},
particles are stochastically accelerated in the IBS
formed by interaction between the neutron star's wind and the companion's wind,
and flow along the contact discontinuity of the two winds.
In addition, some particles can be adiabatically accelerated 
in the flow \citep[e.g.,][]{bkk08}.  Hence, there can be both slow and fast
(accelerated) populations in the shock. Both populations
will emit synchrotron and 
inverse Compton radiation. Because of radiation reaction (competition between acceleration
and energy loss of the emitting particles),
the synchrotron photons emitted by the slow population can only reach energies of $\sim$160\,MeV.
The fast population's bulk Lorentz factor can boost photons to higher energies.
If the off-pulse emission is synchrotron radiation, this might explain the spectral variation,
with the dip phase dominated by slower electrons and the hump including beamed, boosted emission
from the fast population, with a spectral break pushed out above 200\,MeV.

        From the radio eclipses and optical emission lines, we see that the pulsar wind momentum
flux exceeds that of the companion wind. Thus the IBS should wrap around the companion.
In this case we expect a fast population's synchrotron emission to be beamed along
the shock limb, with a phase near $\phi_B \approx 0.25$. In contrast, if the off-pulse 
hump is primarily Compton emission, then we might expect a peak near $\phi_B \approx 0.75$
if optical photons from the heated face of the companion provide the seed population.
This is a somewhat better match to the observed peak near  $\sim 0.8$. Of course a 
more detailed model could account for this asymmetry since
optical emission lines give good evidence that the companion wind is strongly
swept back by orbital motion \citep[e.g.,][]{pp08,rs16}. 

	Unusually for a BW pulsar \citep[e.g.,][]{grmc+14, apg15}
J1311 does not show any clear X-ray orbital modulation in quiescence. This is despite
the fact that, with a photon index $\Gamma\sim1.7$ it seems rather similar to other
BW objects in its average emission. Also, most synchrotron or Compton models of
the gamma-ray emission would predict similar modulation in the lower energy X-ray
population. Perhaps it is then significant that the flare X-ray times appear unevenly
distributed about the orbit. In this case we might assume that these optical/X-ray
events provide the seed photons for gamma-ray Compton emission. Of course some 
portion of the unmodulated X-ray flux might be magnetospheric and a deep
{\it NuSTAR} \citep[][]{hcc+13} study might be able to find pulsations in this component.

	Our connection between the optical and X-ray flares provides some insights
into the nature of these events. In \citet{rfb15}, a flare was observed spectroscopically,
showing that, for that event at least, the optical emission arose from companion
surface and that the energy deposition was well below the photosphere. The event peaked
at phase $\phi_B\approx 0.85$ and covered a substantial fraction $\sim 30$\% of the 
visible surface. The energy flux was two orders of magnitude higher than the pulsar
flux at the orbital radius; we thus infer a rapid release of stored, likely magnetic,
energy. Although this particular event had no X-ray coverage, it would be reasonable to
expect that if the responsible magnetic field has a large coherence scale, penetrating the
surface, then reconnection and particle acceleration above the photosphere could give
rise to synchrotron emission producing the hard $\Gamma\sim1.3$ X-ray events.
If these flares seed the Compton upscattering by IBS-accelerated electrons, gamma-ray
emission at similar phase would result. We note that these flares must not be directly
connected with the pulsar heating since they can also be seen when looking
toward the night face of the companion. A strong, companion-dynamo generated field
might, however, be able to erupt at an any point on the companion surface.
We note that similar correlated optical/X-ray flaring activity is found for
many young stars \citep[e.g.,][]{ftgs07}. These authors find a similar
power-law intensity distribution, but with somewhat smaller indices: 2--2.7. The
average X-ray flare luminosity of the young M stars is typically $\sim 3 \times 10^{29}
{\rm erg\,s^{-1}}$. Using the pulsar DM distance of 1.4\,kpc, the {\it XMM} flare luminosity
and a 10\% duty cycle J1311 has an average flare $L_X \approx 10^{31}{\rm erg\,s^{-1}}$;
we can speculate this higher activity is related to the very short rotation period of
the tidally locked companion star.
Additional flare
studies will be needed to pin down a phase preference, if any, and test this scenario.
Improved, correlated studies with the X-ray events are needed to probe the emission
physics and the extent of the radiating particle energy distribution.
\bigskip

The \textit{Fermi} LAT Collaboration acknowledges generous ongoing support
from a number of agencies and institutes that have supported both the
development and the operation of the LAT as well as scientific data analysis.
These include the National Aeronautics and Space Administration and the
Department of Energy in the United States, the Commissariat \`a l'Energie Atomique
and the Centre National de la Recherche Scientifique / Institut National de Physique
Nucl\'eaire et de Physique des Particules in France, the Agenzia Spaziale Italiana
and the Istituto Nazionale di Fisica Nucleare in Italy, the Ministry of Education,
Culture, Sports, Science and Technology (MEXT), High Energy Accelerator Research
Organization (KEK) and Japan Aerospace Exploration Agency (JAXA) in Japan, and
the K.~A.~Wallenberg Foundation, the Swedish Research Council and the
Swedish National Space Board in Sweden.
 
Additional support for science analysis during the operations phase is gratefully
acknowledged from the Istituto Nazionale di Astrofisica in Italy and the Centre
National d'\'Etudes Spatiales in France. This work performed in part under DOE
Contract DE-AC02-76SF00515.

H.A. acknowledges supports provided by the NASA sponsored {\it Fermi}
Contract NAS5-00147 and by
Kavli Institute for Particle Astrophysics and Cosmology (KIPAC).
This research was supported by Basic Science Research Program through
the National Research Foundation of Korea (NRF)
funded by the Ministry of Science, ICT \& Future Planning (NRF-2017R1C1B2004566).
R.W.R. was supported in part by NASA grant NNX17AL86G.

\bibliographystyle{apj}
\bibliography{GBINARY,BLLacs,PSRBINARY,PWN,STATISTICS,FERMIBASE,COMPUTING}

\end{document}